\documentclass[twocolumn,twocolappendix]{aastex631} 
\usepackage[utf8]{inputenc}
\usepackage{enumerate}
\usepackage{graphicx}
\usepackage{hyperref}
\usepackage{mathtools}
\usepackage{comment}
\usepackage{tabularx}
\usepackage{amsmath}
\usepackage{changepage}
\usepackage{ amssymb }
\usepackage{lipsum} 
\usepackage{float}
\usepackage{commath}
\usepackage{enumerate}
\usepackage{longtable}
\usepackage{stackengine}

\usepackage{cleveref}
\usepackage[caption=false]{subfig}
\usepackage{placeins}
\usepackage[ruled,lined]{algorithm2e}

\AfterEndEnvironment{figure}{\noindent\ignorespaces}

\usepackage{xcolor}

\begin{document}
\title{Slant, Fan, and Narrow: \\  The Response of Stellar Streams to a Tilting Galactic Disk} 
\author[0000-0001-8042-5794]{Jacob Nibauer}
\affiliation{Department of Astrophysical Sciences, Princeton University, 4 Ivy Ln, Princeton, NJ 08544, USA}

\author[0000-0002-7846-9787]{Ana Bonaca}
\affiliation{The Observatories of the Carnegie Institution for Science, 813 Santa Barbara Street, Pasadena, CA 91101, USA}

\author[0000-0002-8495-8659]{Mariangela Lisanti}
\affiliation{Department of Physics, Princeton University, Princeton, NJ 08544, USA}
\affiliation{Center for Computational Astrophysics, Flatiron Institute, New York, NY 10010, USA}

\author[0000-0002-8448-5505]{Denis Erkal}
\affiliation{Department of Physics, University of Surrey, Guildford, GU2 7XH, UK}

\author{Zoe Hastings}
\affiliation{Department of Physics, University of Surrey, Guildford, GU2 7XH, UK}

\correspondingauthor{Jacob Nibauer}
\email{jnibauer@princeton.edu}

\begin{abstract}
\noindent
Stellar streams are sensitive tracers of the gravitational potential, which is typically assumed to be static in the inner Galaxy.
However, massive mergers like Gaia-Sausage-Enceladus can impart torques on the stellar disk of the Milky Way that result in the disk tilting at rates of up to 10--20\,deg\,Gyr$^{-1}$. Here, we demonstrate the effects of disk tilting on the morphology and kinematics of stellar streams. Through a series of numerical experiments, we find that streams with nearby apocenters $(r_{\rm apo} \lesssim 20~\rm{kpc}$) are sensitive to disk tilting, with the primary effect being changes to the stream's on-sky track and width. Interestingly, disk tilting can produce both more diffuse streams and more narrow streams, depending on the orbital inclination of the progenitor and the direction in which the disk is tilting. Our model of Pal 5's tidal tails for a tilting rate of $15\,\rm{deg}\,\rm{Gyr}^{-1}$ is in excellent agreement with the observed stream's track and width, and reproduces the extreme narrowing of the trailing tail. We also find that failure to account for a tilting disk can bias constraints on shape parameters of the Milky Way's local dark matter distribution at the level of $5$--$10\%$, with the direction of the bias changing for different streams. Disk tilting could therefore explain discrepancies in the Milky Way's dark matter halo shape inferred using different streams.
 
\vspace{1cm}
\end{abstract}

\section{Introduction}
Under the standard $\Lambda$ cold dark matter ($\Lambda$CDM) paradigm, the growth of galaxies like the Milky Way is governed by mergers with smaller satellites \citep{1978MNRAS.183..341W,1995MNRAS.275...56N}. The dynamical effect of a major merger on the baryonic structure of a galaxy is potentially substantial, since the total mass of a satellite can be comparable to, or greater than, the stellar mass of the host. The accretion of a massive satellite can lead to several effects on the host galaxy's baryonic and dark matter components, including disk warping \citep{2020NatAs...4..590P,2023ApJ...957L..24H}, a net reflex motion of the galactic disk \citep{2021MNRAS.506.2677E,2021NatAs...5..251P}, phase spirals \citep{2018Natur.561..360A}, and wakes in the stellar halo \citep{2019MNRAS.488L..47B,2021Natur.592..534C} and the dark matter halo \citep{2019ApJ...884...51G}. In the Milky Way, there is evidence of past and ongoing mergers with (e.g.) the Magellanic clouds, Sagittarius, and the Gaia Sausage Enceladus (GSE) \citep{2018MNRAS.478..611B,2018Natur.563...85H,2018MNRAS.481..286L,2021Natur.592..534C}. These interactions challenge the validity of static, equilibrium models for the Galaxy and motivate the use of tracer populations to infer the time-evolution of the Milky Way. 

In this work, we focus on one such time-dependent effect predicted in $\Lambda\rm{CDM}$: stellar disk-tiling \citep{1986MNRAS.218..743B, 1989MNRAS.237..785O,1997ApJ...480..503H}. Disk tilting is a natural consequence of a major merger with a net angular momentum vector that differs from that of the stellar disk. As the satellite's orbit decays, there is a transfer of angular momentum from the infalling satellite to the host's disk \citep{1997ApJ...480..503H}. Unlike disk warping, disk tilting is only relevant for the inner regions of a stellar disk for which the stellar mass is tightly coupled due to self gravity. In these regions, several works have shown that the disk behaves as a nearly rigid body in response to an infalling satellite \citep{1986MNRAS.218..743B,1998ApJ...506..590S,2006MNRAS.370....2S,2023MNRAS.518.2870D}. The rigidity of the inner disk can lead to substantial changes in its orientation over time in response to external torques \citep{1999MNRAS.304..254V,2004MNRAS.351.1215B,Helmi, 2009ApJ...700.1896K,2017MNRAS.469.4095E,2017MNRAS.472.3722G,2022MNRAS.513.1867D,2022MNRAS.510.1375S,2023MNRAS.518.2870D,2023arXiv231013050C}. Other sources of external torques include halo triaxality, or dark matter figure rotation, both of which can induce disk tilting even in the absence of infalling satellites \citep{2012MNRAS.426..983D,2015MNRAS.452.2367Y,2015MNRAS.452.4094D}. The accretion of gas onto the disk provides another external torque which can further drive disk tilting \citep{2015MNRAS.452.4094D,2019MNRAS.488.5728E}. 

Typical rates for disk tilting in simulations vary significantly depending on the source of external torque, the merger history of the host galaxy, and the nature of the simulation considered (i.e., $N-$body-only versus hydrodynamical). Through a series of zoom-in hydrodynamical simulations, \citet{2017MNRAS.469.4095E} find a change in the disk's angular momentum vector by as much as $\sim 25~\rm{deg} / \rm{Gyr}$ when measured from $z = 0.3$ to the present day. \citet{2023MNRAS.518.2870D} find average tilting rates at the level of $\sim 10$--$15~\rm{deg}/\rm{Gyr}$ for a GSE-like merger, with the disk's tilt continuing even after the dissolution of the satellite. This is due to the accretion of the satellite's dark matter onto the inner galaxy, whose aspherical distribution function can produce so called ``tilting modes" \citep{1997ApJ...480..503H,1998ApJ...506..590S} in the galactic disk. In the high-resolution ARTEMIS cosmological simulations, \citet{2022MNRAS.513.1867D} find disk tilting rates of up to a remarkable $60~\rm{deg}/\rm{Gyr}$, with lower tilting rates ($\sim 10~\rm{deg}/\rm{Gyr}$) sustained for several Gyrs. The ability of external torques to sculpt the stellar disk both morphologically and dynamically is an expected outcome of galaxy formation in a $\Lambda\rm{CDM}$ cosmology, though there is a large amount of variance in the expected magnitude of stellar disk tilting.

Considering the Milky Way's rich accretion history, we may expect the stellar disk of the Galaxy to be tilting in response to external torques. There is currently no observational constraint on the tilting rate of the Milky Way's disk. One promising observable of disk tilting is proposed in \citet{2014ApJ...789..166P}: by calibrating kinematic measurements of disk stars to a reference frame defined by distant quasars, they predict a relative proper motion accuracy of greater than $1\mu\rm{as}~\rm{yr^{-1}} = 0.28~\rm{deg}/\rm{Gyr}$ in the quasar frame using {\it Gaia} astrometry. This is well below disk tilting rates predicted from cosmological simulations, providing a promising avenue to detect coherent motion of the stellar disk in excess of the usual galactic rotation. Still, this measurement is complicated by the acceleration of the solar system's barycenter with respect to the Galactic center. While such a measurement has been made using distant quasars (see \citealt{2021A&A...649A...9G}), it is worth considering other observables associated with a tilting disk.

In this paper, we explore whether kinematically cold, globular cluster stellar streams can provide a set of observables to constrain stellar disk tilting. We focus on stellar streams since they are sensitive probes of the gravitational potential and have been used extensively to constrain static models of the Milky Way (e.g., \citealt{2010ApJ...712..260K,2016ApJ...833...31B,2019MNRAS.486.2995M}). In addition, stellar streams are also sensitive to time-dependent perturbations, such as the stellar bar \citep[e.g.][]{2016MNRAS.460..497H,2017MNRAS.470...60E,2017NatAs...1..633P}, the Large Magellanic Cloud \citep[e.g.][]{2013ApJ...773L...4V,2019MNRAS.487.2685E,2021ApJ...923..149S,2023MNRAS.521.4936K}, the deformations of the Milky Way and Large Magellanic Cloud in response to each other \citep{2023MNRAS.518..774L}, dark matter figure rotation \citep{2021ApJ...910..150V}, and dark matter subhalos \citep{2002MNRAS.332..915I,2002ApJ...570..656J,2012ApJ...748...20C,2016MNRAS.463..102E,2019ApJ...880...38B,2021MNRAS.502.2364B}. Here we construct a series of numerical experiments to measure the effect of a tilting disk on stellar streams. We also develop a perturbative model to make predictions for globular cluster streams.

The paper is organized as follows. In \S\ref{sec: simulations}, we discuss our simulation procedure. In \S\ref{sec: effect_tilting_disk_streams}, we demonstrate the kinematic and morphological effects of a tilting disk on streams. In \S\ref{sec: effect_stream_width}, we model the angular momentum evolution of streams with a tilting disk. In \S\ref{sec: sensitivity}, we consider the effect of a tilting disk on Milky Way globular cluster streams and compare to observations of Pal 5. Implications for potential reconstruction are also explored. We discuss our results in \S\ref{sec: discussion} and conclude in \S\ref{sec: conclude}.

\section{Simulations}\label{sec: simulations}
In this section, we describe the details of our simulations, including our implementation of a time-dependent disk potential and the parameters of the stream models considered.

\subsection{Coordinate System}\label{sec: coords}
An illustration of the disk-aligned, co-rotating coordinate system is shown in Fig.~\ref{fig: coord_sys}. The primed coordinates ($x^\prime, y^\prime, z^\prime$) are aligned with the disk at all times and coincide with the inertial frame (represented with unprimed coordinates: $x,y,z$) at the present day. We typically view all of our simulations at the present day, so that a distinction does not have to be drawn between co-rotating spatial coordinates and inertial coordinates. 

For a stream progenitor, the orbital inclination today is given by the angle $\psi$ between the $z^\prime$-axis and the angular momentum vector ($\mathbf{L}$) of the orbit. The azimuthal angle between the angular momentum vector and the $x^\prime-$axis is $\phi$. We will also discuss the azimuthal angle, $\alpha$, between the tilting axis of the disk, $\hat{\mathbf{n}}_{\rm tilt}$, and the angular momentum vector $\mathbf{L}$. Finally, the inclination of the tilting axis today is given by the angle $\beta$ between the $z^\prime$-axis and $\hat{\mathbf{n}}_{\rm tilt}$.

Throughout the paper, we reference the effect of precession and nutation. The former corresponds to azimuthal evolution of $\mathbf{L}$ (i.e., in the $\hat{\phi}$ direction), while the latter corresponds to oscillations of $\mathbf{L}$ in $\hat{\psi}$. The effect of precession and nutation on streams in a static potential is characterized in \citet{2016MNRAS.461.1590E}.

\subsection{Potentials}\label{sec: potentials}
In this section we specify the fiducial potential models used throughout the paper. We consider a simple galaxy model consisting of a spherical Navarro-Frenk-White~(NFW) halo \citep{1996ApJ...462..563N} and a Miyamoto-Nagai disk \citep{1975PASJ...27..533M}. The NFW halo has density 
\begin{equation}
    \rho_{\rm halo}\left(x,y,z\right) = \frac{M_{\rm halo}}{4\pi r_s^3}\frac{1}{\left(m/r_s\right)\left(1 + m/r_s \right)^2},
\end{equation}
where 
\begin{equation}
    m^2 \equiv x^2 + y^2 + \frac{z^2}{q^2_{z,\rm{halo}}}.
\end{equation}
For the majority of this work, we assume a spherical halo with $q_{z,\rm{halo}} = 1$. We adopt a scale radius of $ r_s = 15~\rm{kpc}$ and choose the mass of the halo ($M_{\rm halo}$) so that the circular velocity at $8~\rm{kpc}$ is $\approx 240~\rm{km/s}$ when accounting for both the disk (to be discussed) and halo. These choices are not meant to match the Milky Way precisely but are similar to constraints from \citet{2019MNRAS.486.2995M,2023MNRAS.521.4936K}, derived from observations of Milky Way streams.

The Miyamoto-Nagai disk potential has the functional form
\begin{equation}
    \Phi_{\rm disk}\left(x^\prime,y^\prime,z^\prime\right) = \frac{-GM_{\rm Disk}}{\sqrt{x^{\prime 2} + y^{\prime 2}  + \left( \sqrt{z^{\prime 2} + b^2} + a\right)^2 }},
\end{equation}
where we have used primed coordinates that are aligned with the disk. We set $M_{\rm Disk} = 6.8\times10^{10}M_\odot$, $a = 3~\rm{kpc}$, and $b = 0.28~\rm{kpc}$. These values come from \texttt{MWPotential2014} in \cite{2015ApJS..216...29B} and are similar to constraints for the Milky Way's disk (e.g., \citealt{2013ApJ...779..115B}).

We treat the disk as a rigid body, such that it tilts around an axis $\hat{\mathbf{n}}_{\rm tilt}$ passing through its center of mass with frequency $\Omega_{\rm Disk}$. This rigid body treatment is an idealization, though several authors have argued that at least the inner regions of the disk behave rigidly in response to external torques \citep{1986MNRAS.218..743B,1998ApJ...506..590S,2006MNRAS.370....2S,2023MNRAS.518.2870D}. Therefore, our treatment likely reveals the first-order effect of a tilting disk on streams, though future work could consider more realistic simulations.

We implement the disk's tilt through a time-dependent rotation matrix $\mathbf{R}\left(\Omega_{\rm Disk} t; \hat{\mathbf{n}}_{\rm tilt} \right)$ that maps from the inertial $(x,y,z)$ frame to the primed $(x^\prime, y^\prime, z^\prime)$ frame. We have ensured that our implementation produces the same sets of orbits as one would find by integrating the equations of motion in a rotating frame. Our implementation allows us to flexibly extend the tilting disk potential to other scenarios where we would like (e.g.) an aspherical halo component to remain stationary as the disk tilts. The tilting disk potential is implemented as a \texttt{Gala} compatible C extension on github.\footnote{\url{https://github.com/jnibauer/TiltingDiskPotential}}

\subsection{Orbit Integration and Stream Generation}

\begin{figure}
    \centering\includegraphics[scale=.5]{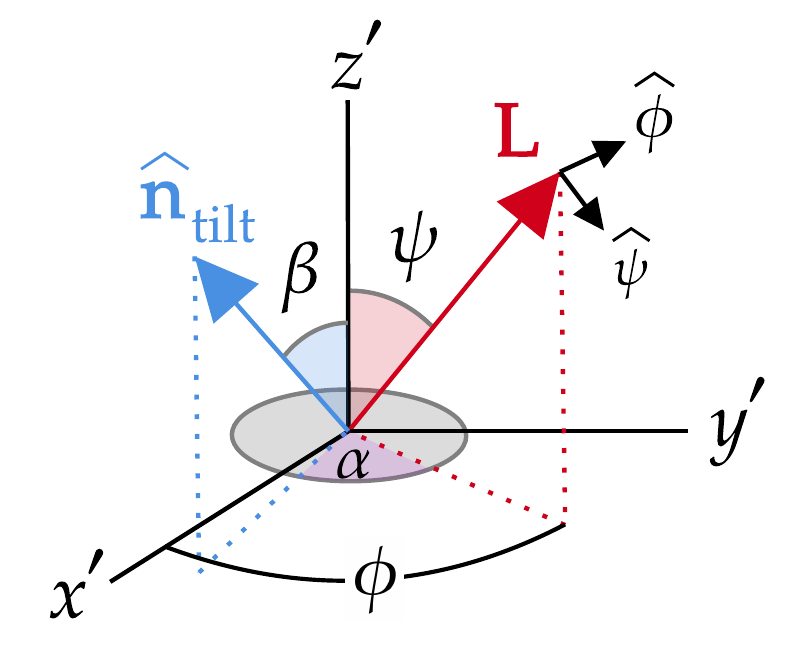}
    \caption{Illustration of the coordinate frame referenced throughout the paper. This disk is shown in gray, with the orbital angular momentum vector, $\mathbf{L}$, shown in red. The orbital inclination is $\psi$, and the tilting axis of the disk is $\hat{\mathbf{n}}_{\rm tilt}$. In this work, we consider the specific case with $\beta = 90~\rm{deg}$.}
    \label{fig: coord_sys}
\end{figure}

For generating streams, we utilize the particle-spray method \citep{2015MNRAS.452..301F} implemented in \texttt{Gala} \citep{gala}. This method has been shown to reproduce streams in $N-$body simulations. We consider kinematically cold globular cluster streams because of their well-localized spatial and kinematic properties, which make them sensitive tracers of a galaxy's mass distribution. We consider a fiducial progenitor with mass $3\times10^4 M_\odot$ and a scale radius of $4~\rm{pc}$. We evolve each stream for $4~\rm{Gyr}$ unless otherwise noted. The location of the stream progenitor today is specified by its instantaneous orbital plane (given by its angular momentum vector), its distance from the galactic center, and its azimuthal position within the orbital plane. The speed of the progenitor is set to a fraction of the local circular speed, $f v_{\rm circ}$, where $f \in [0.5,1]$. The direction of the progenitor's velocity vector is orthogonal to the progenitor's position vector, $\mathbf{r}$. We parameterize the orbit this way because the effect of a tilting disk is sensitive to the inclination, and therefore angular momentum vector, of an orbit.

We will typically compare streams generated in a static potential to those generated in a time-dependent tilting disk potential. In most cases, the final (i.e., observed) phase-space location of the progenitor is fixed to be the same for the two scenarios.

\subsection{Stream Frame and Track Fitting}\label{sec: frame_and_splines}
Throughout the paper, we fit several streams with splines to provide summary statistics for the effect of a tilting disk on stream morphology and kinematics. We briefly describe our spline fitting procedure below.

We view each stream in its great circle frame, characterized by two on-sky angles $\phi_1,\phi_2$, two proper motions $\mu_{\phi_1}, \mu_{\phi_2}$, distance $r$, and radial velocity $v_r$. The angle $\phi_1$ is along the elongated axis of each stream, allowing us to parametrize each of the other dimensions listed above as a function of $\phi_1$. We fit splines to the mean ``track" in each dimension of the great circle frame, as a function of $\phi_1$. We utilize univarite splines of degree $k = 5$ implemented in \texttt{Scipy} \citep{2020SciPy-NMeth}. To estimate stream width, we compute the standard deviation in $\phi_2$ of the stream particles in $\sim$30 bins of $\phi_1$ taken over the length of the stream. The minimum and maximum $\phi_1$ values are set by determining the 80\% highest density $\phi_1$ interval for the static potential stream. This ensures that the underpopulated regions of the tidal tails are excluded in our fit and analysis of each stream.

 \section{The Effect of a Tilting Disk on Streams}\label{sec: effect_tilting_disk_streams}
 In this section, we illustrate the effect of realistic disk tilting rates on the morphology and kinematics of stellar streams. 

 \begin{figure*}
    \centering
    \includegraphics[scale=0.48]{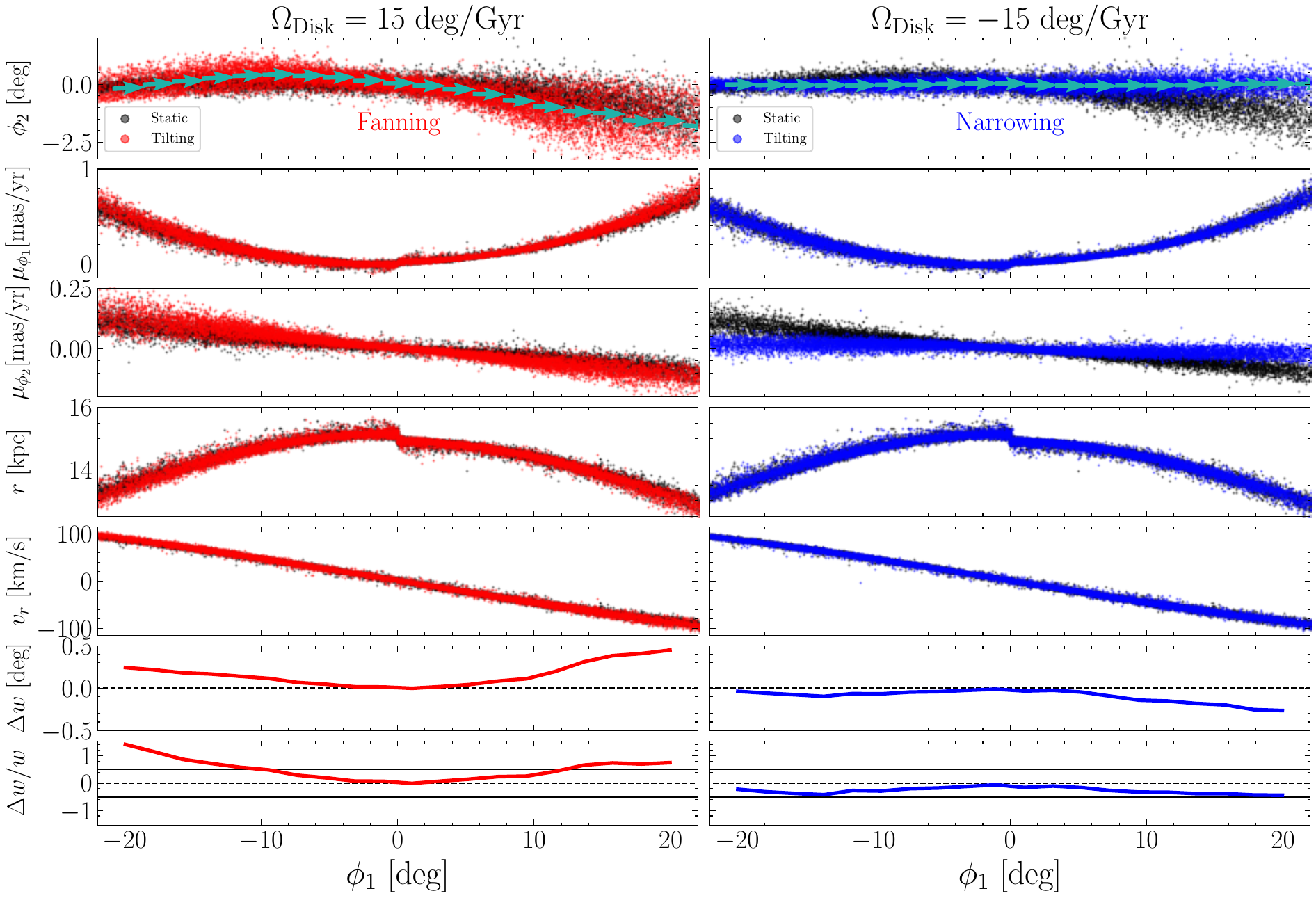}
    \caption{Three typical streams are illustrated in the same (inertial) galactocentric great circle frame. The black stream in both panels is generated in a static potential, while the red and blue streams are generated in a tilting disk potential with rotation around $+\mathbf{x}$ (red) and $-\mathbf{x}$ (blue).  As a result of the tilting disk, the orbital plane of the red stream was more aligned with the disk's midplane in the past, while the blue stream's orbital plane was comparatively misaligned. The final progenitor location is the same for all cases. Cyan arrows depict the average proper motion vectors along the red and blue streams. The bottom two rows in both columns represent the difference in stream widths and the fractional difference relative to the unperturbed model. Solid lines in the bottom row represent a change in stream width of $\pm 50\%$.
    The red stream appears to fan out, while the blue stream is more narrow. }
    \label{fig: HeatingCoolingStream}
\end{figure*} 

\subsection{Effect on Stream Morphology and Kinematics}\label{sec: effect_on_morphology_and_kinematics}
We carry out a grid of around 1000 simulations with different stream orbits and disk-tilting rates. To illustrate the effect of a tilting disk on streams, we first show a few typical streams from our simulations and then present a summarized viewed of the simulation grid. 

Three typical streams from our simulations are shown in Fig.~\ref{fig: HeatingCoolingStream}, viewed in an inertial galactocentric great circle frame. In both columns the black stream is the same, and is generated in a static potential. The red and blue streams are generated in a tilting disk potential with rotation around $+\mathbf{x}$ and $-\mathbf{x}$, respectively. Cyan arrows represent mean proper motion vectors for the red and blue streams. The tilting rate is $15~\rm{deg}/\rm{Gyr}$. All three streams have their progenitor at the same final phase-space location at a distance of $r = 15~\rm{kpc}$ with velocities equal to 60\% of the local circular velocity. The inclination of the progenitor's orbit at the final snapshot is $\psi = 70~\rm{deg}$, and the azimuthal location of the angular momentum vector is $\phi = 40~\rm{deg}$. In the fiducial static potential, the progenitor has a pericenter of $\sim 6.4~\rm{kpc}$ and an apocenter of $\sim 15~\rm{kpc}$. Because the disk always ends up with the same orientation at the present day for all simulations, the $+\hat{\mathbf{x}}$ ($-\hat{\mathbf{x}}$) rotation aligns (misaligns) the disk with the present-day orbital plane as a function of increasing lookback time.

The most significant differences between the streams generated in a static potential compared to the tilting disk potential are ($i$) the track of the streams are offset by $\sim 1$--$2~\rm{deg}$ in $\phi_2$, and ($ii$) the width of the streams differ by $\sim 0.3$--$0.4~\rm{deg}$ in the extended tidal tails. The other kinematic dimensions differ only slightly compared to the stream generated in a static potential.

The direction of the disk's tilt ($\pm{\mathbf{x}}$) is important in determining the width of a stream. For instance, the red stream in Fig.~\ref{fig: HeatingCoolingStream} systematically fans out over its length, while the blue stream systematically narrows compared to the static model stream. The fractional difference in stream widths relative to the static model is shown in the bottom row of Fig.~\ref{fig: HeatingCoolingStream}. The red stream experiences a substantial amount of fanning, with a fractional increase in its width of over $100\%$ compared to the static model stream (for $\phi_1 < 0~\rm{deg}$). The blue stream experiences substantial narrowing and is $\sim 50\%$ more narrow than the stream generated in the static potential. The divergence of stream widths---yielding both thinner and thicker streams---is a common result for progenitors on more inclined orbits relative to the disk. This suggests that stream width is one of the key stream-based observables of a tilting disk. We discuss the dynamical origin of this effect in \S\ref{sec: effect_stream_width}. 

In \citet{2019MNRAS.487.2685E}, it is shown that a clear observable of a time-dependent potential is the misalignment between the track of a stream and its (reflex-corrected) proper motions. We find that a tilting disk potential, at least with $\Omega_{\rm Disk}$ in the range predicted by cosmological simulations, does not produce this misalignment. This is reflected by the alignment between the cyan vectors (proper motions) in Fig.~\ref{fig: HeatingCoolingStream} and the stream track. We discuss the lack of a misalignment signal further in \S\ref{sec: tilting_orbital_poles}.

\begin{figure*}
    \centering
    \includegraphics[scale=0.46]{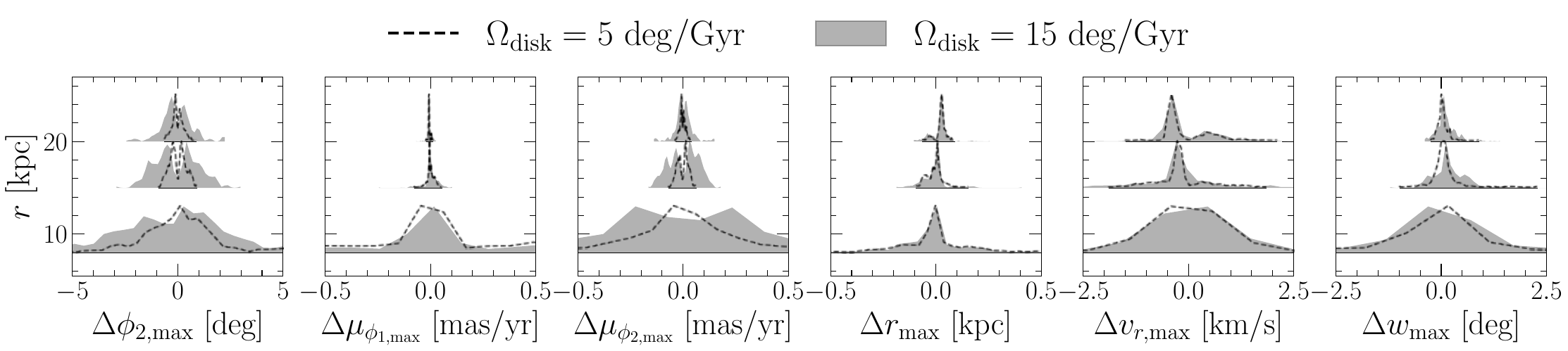}
    \caption{From a grid of $\sim 1000$ simulations, we illustrate the distribution of maximum absolute differences in stream morphology and kinematics between streams in the tilting disk potential and the corresponding stream evolved in a static potential. $\Delta$ corresponds to the difference ``$\rm{perturbed} - \rm{static}$". The maximum difference is taken over the length of each stream. The distributions for each observable are shown as a function of the stream's distance from the galactic center, $r \in \{8,15,20\}~\rm{kpc}$. Dashed distributions correspond to a $5~\rm{deg}/\rm{Gyr}$ tilt, while the solid distribution correspond to $15~\rm{deg}/\rm{Gyr}$. We show the distribution of maximum phase-space differences for different apocentric radii. The rightmost panel corresponds to the maximum (in absolute terms) width difference between the perturbed and static stream model. Streams with closer-in apocenters experience the most substantial perturbations due to disk-tilting. Even small amounts of disk-tiling can produce observable changes in the kinematics and morphology of a stream.} 
    \label{fig: summary_stream_plot}
\end{figure*}

In Fig.~\ref{fig: summary_stream_plot}, we provide a summarized view of $\sim 1000$ stellar streams generated over a grid of orbits in the tilting disk potential and static potential. Our grid of streams includes progenitors with final inclinations $\psi \in [0,90]~\rm{deg}$, distributed over the full range of azimuthal angles $\phi \in [0,360)~\rm{deg}$. The progenitor orbits range from somewhat radial ($f =0.6$) to approximately circular ($f=1$). We consider streams at three distances $r \in \{8,15,20\}~\rm{kpc}$ subject to a tilting disk with $\Omega_{\rm Disk} \in \{5, 15\}~\rm{deg}\rm{/Gyr}$ and $\hat{\mathbf{n}}_{\rm tilt } = \hat{\mathbf{x}}$. All streams are evolved for $4~\rm{Gyr}$.

To compress the large grid of simulations down to an interpretable set of summary statistics, we view each stream in its great circle frame and fit its mean on-sky track and kinematics using a series of splines. The spline fitting procedure is outlined in \S\ref{sec: frame_and_splines}. The difference between the spline tracks in the static potential and tilting disk potential (as a function of $\phi_1$) is computed and the distribution of maximum absolute differences taken over the length of each stream is shown in Fig.~\ref{fig: summary_stream_plot} as a function of  apocentric radius. Spline tracks are fit at apocenter. The $\Delta$ symbol in the label of each panel represents a difference between the perturbed model and static model (i.e., $\Delta$ corresponds to ``perturbed - static"). 

Fig.~\ref{fig: summary_stream_plot} can be interpreted as showing the typical order-of-magnitude effect on streams in each of the observable dimensions. Unsurprisingly, the extent of each distribution in Fig.~\ref{fig: summary_stream_plot} diminishes with increasing distance. This is because streams with more distant apocenters are less sensitive to the effects of a tilting disk. Additionally, a higher value of $\Omega_{\rm Disk}$ typically produces broader distributions in Fig.~\ref{fig: summary_stream_plot}, or larger deviations from the static model. The largest observable difference is in the on-sky location of each stream, with typical offsets in $\phi_2$ at level of $1~\rm{deg}$ or more. This is substantial, considering that the angular position uncertainty from (e.g.) {\it Gaia} measurements are at the milliarcsecond level. Fractional changes in proper motions, distances, and radial velocities are typically small, especially for streams with more distant apocenters. 
The rightmost panel of Fig.~\ref{fig: summary_stream_plot} is perhaps the most intriguing: stellar disk-tilting can produce large differences in stream width. Namely, the tails of the $\Delta w_{\rm max}$ distribution indicate that disk-tilting can produce both more diffuse streams ($\Delta w_{\rm max} > 0$) and more narrow streams ($\Delta w_{\rm max} < 0$).

\begin{figure*}
    \centering
    \includegraphics[scale=0.6]{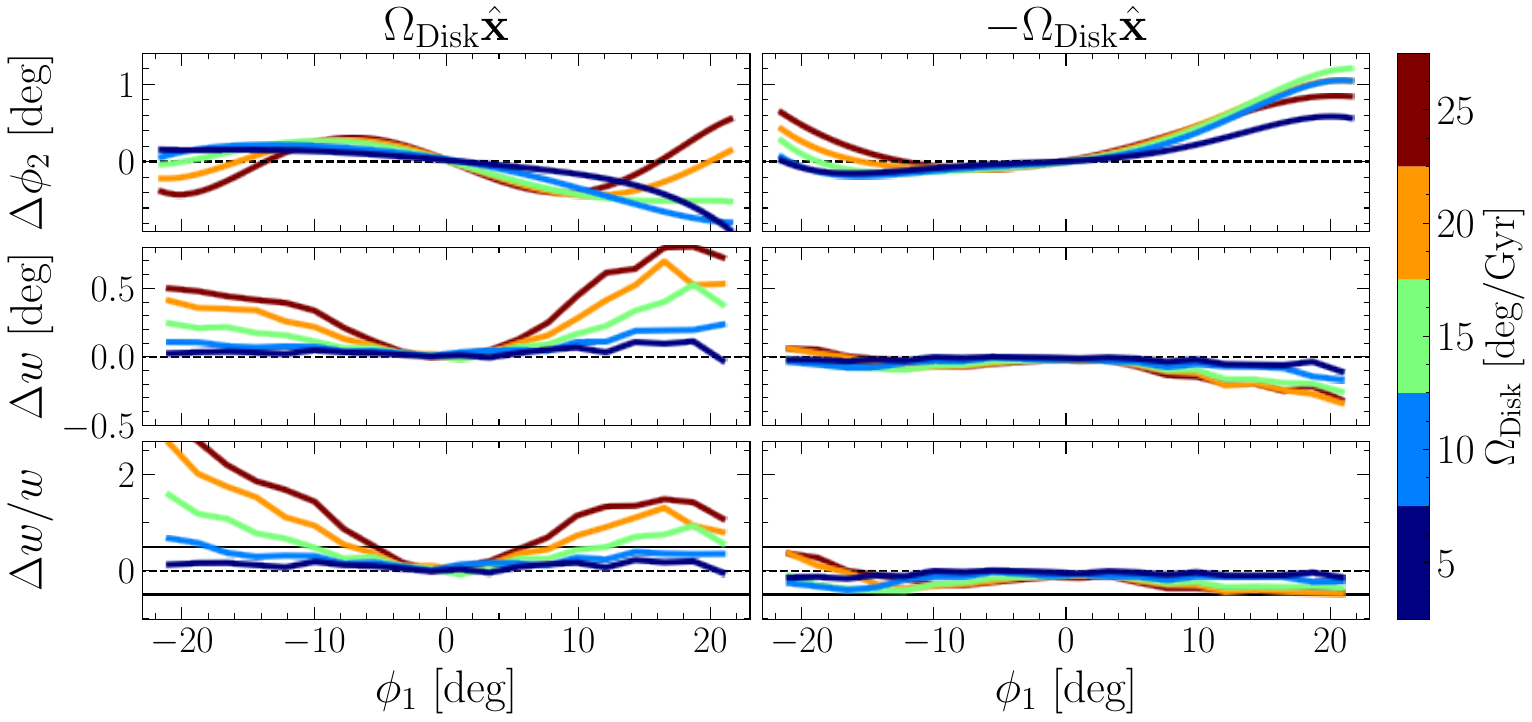}
    \caption{The effect of varying the tilting rate on the track and width of a typical stream in our simulations (same final progenitor as in Fig.~\ref{fig: HeatingCoolingStream}). In the left column, rotation around $ +\mathbf{x}$ leads to systematically more diffuse streams ($\Delta w > 0$) and large $\phi_2$ offsets relative to the unperturbed model. In the right column, rotation around the opposite direction, $-\mathbf{x}$, leads to more narrow streams ($\Delta w < 0$ typically) and $\phi_2$ offsets in the opposite direction. The solid lines in the bottom row represent a change in stream width of $\pm 50\%$. All tilting rates produce detectable differences in stream morphology. }
    \label{fig: VaryOmegaDisk_summary}
\end{figure*} 

\subsection{Varying the Tilting Rate}
We now illustrate the effect of varying the magnitude of $\mathbf{\Omega}_{\rm disk}$ on the track and width of streams. 

We generate streams for five values of $\Omega_{\rm Disk} \in [5,25]~\rm{deg}/\rm{Gyr}$ with the same final progenitor location as in Fig.~\ref{fig: HeatingCoolingStream}. For a given value of $\Omega_{\rm Disk}$, we tilt the disk around both $+{\mathbf{x}}$ and $-{\mathbf{x}}$. Once a stream has been generated, we convert to its great circle galactocentric stream frame and fit the mean stream track with splines using the procedure outlined in \S\ref{sec: frame_and_splines}. 

The result of this experiment is illustrated in Fig.~\ref{fig: VaryOmegaDisk_summary}. The first column corresponds to the $+\mathbf{x}$ rotation, while the second represents the opposite rotation. The first row shows the difference in stream tracks taken relative to the corresponding stream generated in a static potential model (i.e., $\Delta \phi_2 = \phi_{2,\rm{perturbed}} - \phi_{2,\rm{static}}$), while the second row shows the difference in each stream's width. The third row shows the fractional change in stream width. For both tilting directions, increasing the value of $\Omega_{\rm Disk}$  typically produces larger offsets in $\phi_2$. Even for small to moderate values of $\Omega_{\rm Disk} \in [5,10]~\rm{deg}/\rm{Gyr}$, the offsets are easily measurable, with $\Delta \phi_2 \gtrsim 0.5~\rm{deg}$.

Fig.~\ref{fig: VaryOmegaDisk_summary} (second and third rows) also illustrates changes in the width of streams generated in the tilting disk potential, as a function of different tilting rates. For streams in the left column, the typical difference in stream width is $\sim 0.2~\rm{deg}$. Unsurprisingly, streams in the left column of Fig.~\ref{fig: VaryOmegaDisk_summary} with the $+\mathbf{x}$ rotation tend to fan out more with higher tilting rates. The fractional change in the stream width compared to the corresponding stream in the static potential is $> 100\%$ for $\Omega_{\rm Disk} \sim 15~\rm{deg}/\rm{Gyr}$ and higher. This indicates that the perpendicular extent of a stream (orthogonal to the elongated stream track) should provide a useful diagnostic of global time-dependence in the gravitational potential. Even a $5~\rm{deg}/\rm{Gyr}$ tilt produces width variations at the level of $\sim 0.1~\rm{deg}$, which is detectable with current astrometric and photometric capabilities from (e.g.) {\it Gaia}, DECam, and the Southern Stellar Stream Spectroscopic Survey (S$^5$) (e.g., \citealt{2020MNRAS.493.4978S,2020ApJ...889...70B,2021ApJ...923..149S}).

The track offsets and width differences for streams generated with the same final progenitor location but with the disk tilting around the opposite direction ($-{\mathbf{x}}$) are shown in the right column of Fig.~\ref{fig: VaryOmegaDisk_summary}. First, we note that the offset in $\phi_2$ for this set of streams has a comparable magnitude but opposite sign relative to the left column. The behavior in stream width for the two tilting directions is also notably different. While the width always increases for the $+\mathbf{x}$ rotation ($\Delta w$ is strictly positive for rotation about $+{\mathbf{x}}$), rotation around $-{\mathbf{x}}$ easily produces streams that appear thinner ($\Delta w < 0$) than the equivalent stream in a static potential. For realistic values of $\Omega_{\rm Disk}$, we find a roughly 50\% reduction in the width of the stream compared to its static potential counterpart. 
The substantial fanning and narrowing of streams in this scenario provides a promising observational outlook for not only detecting disk tilting, but also in localizing the disk's tilting direction according to which streams appear thinner or thicker. We discuss the dynamical origin of this behavior in \S\ref{sec: effect_stream_width}.

\subsection{Time-Dependence of $L_z$ and Tilting of Orbital Poles}\label{sec: tilting_orbital_poles}
We now consider the orbital pole distribution of streams in the presence of a tilting disk, which is related to the angular momentum evolution of a stream. A full measurement of a stream's orbital poles requires 6D phase-space observations, but does not require knowledge of the potential. 

In a static axisymmetric potential, angular momentum about the symmetry axis ($L_z$) is conserved. The conservation of $L_z$ demands that motion is confined to the 2D meridional plane spanned by $(R,z)$, which rotates around the symmetry axis. For a spherical halo with a static axisymmetric disk, $\dot{L}_z = 0$ while the other components of the angular momentum vector, $(L_x, L_y)$, tend to oscillate due to torques applied by the disk. This gives rise to the precession and nutation of orbital poles described in \S\ref{sec: coords}.

\begin{figure*}
    \centering\includegraphics[scale=0.55]{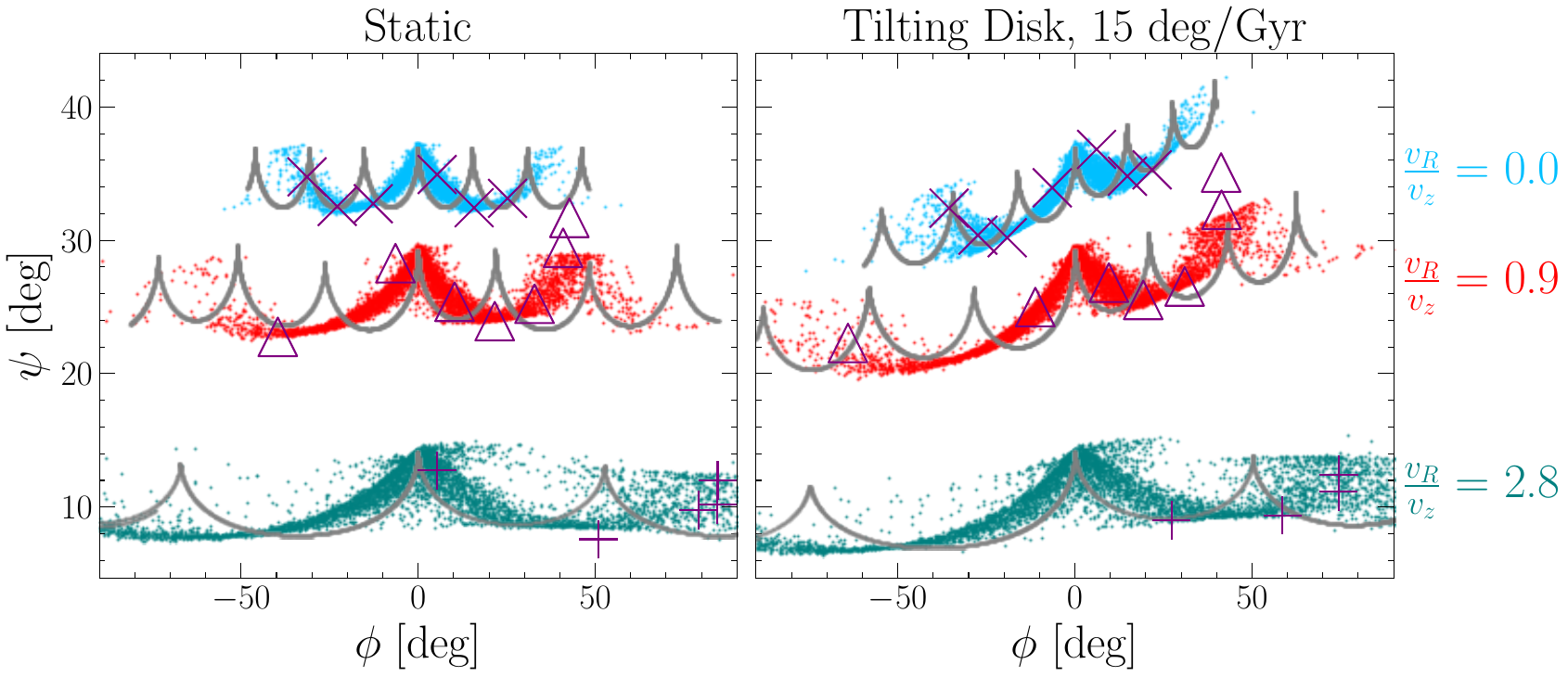}
    \caption{Comparison between the orbital pole distribution of streams and their progenitors in the fiducial static potential (left) and tilting disk potential (right). We generate three streams in each case from the same orbit family, with varying vertical excursions parameterized through the velocity ratio $v_R/v_z$. Each stream progenitor share the same final spatial location, but different orbital pole distributions. In both columns, the gray curves represent the progenitor's orbital pole evolution integrated for $\pm 250~\rm{Myr}$ from the present day. In the right panel, the orbital pole evolution of the progenitor is calculated in a frame co-rotating with the disk (so $\psi$ is always the angle between the disk's normal and the orbital angular momentum). In both panels, plus symbols, triangles, and crosses represent the normal vectors of planes fit to the \emph{spatial} position of stream segments, split into $20~\rm{deg}$ arcs. In the static and tilting disk potentials, the stream's pole vectors, progenitor's pole evolution, and spatial plane of each stream are closely aligned. There is no significant track-velocity misalignment even for a moderate disk-tilting rate of $15~\rm{deg}/\rm{Gyr}$.}
    \label{fig: StreamAngularMomentumEvo}
\end{figure*} 

The left panel of Fig.~\ref{fig: StreamAngularMomentumEvo} provides an illustration of the orbital pole distribution for three streams generated in the fiducial static potential. Each stream progenitor is drawn from the same orbit family, with a common energy and $L_z$, though different vertical motions parameterized by the ratio $v_R/v_z$. Each progenitor ends up at the same spatial location after 5~Gyr, $(x,y,z) = (15,0,0)~\rm{kpc}$, though with different velocities according to the above recipe. The azimuthal location of each progenitor's orbital angular momentum vector is $\phi = 0~\rm{deg}$, while the inclination ($\psi$) varies as determined by the ratio $v_R/v_z$.

The time evolution of orbital pole vectors for each progenitor is shown in gray, integrated for $\pm 250~\rm{Myr}$ from the present day. In a static axisymmetric potential, \citet{2016MNRAS.461.1590E} showed that each stream's local spatial plane coincides with ($i$) the time-evolving orbital plane of the stream progenitor and ($ii$) the stream's local orbital plane. We now demonstrate this result from \citet{2016MNRAS.461.1590E} by comparing the spatial plane of each stream to the distribution of its orbital poles. We then consider whether the same behavior holds for a tilting disk potential. 

We divide each stream into $20~\rm{deg}$ segments and fit planes to the particles in position space. The resulting normal vectors to the planes are shown as the purple symbols. First, we remark that the static potential gives rise to precession of the progenitor's orbital poles (evolution in azimuth; $\phi$) as well as nutation (small oscillations in elevation; $\psi$). As expected from \citet{2016MNRAS.461.1590E}, each stream follows the orbital pole evolution of its progenitor both in its kinematic and spatial location. Neglecting the oscillations in $\psi$ due to nutation (which are subdominant to the azimuthal evolution due to precession), the orbital pole distribution of each stream can be thought of as a cone tracing a constant $\psi$ section. In the static case, there is some misalignment between the purple markers and the scatter-points. This is because kinematic oscillations (scatter-points) precede spatial oscillations of the stream (purple markers). Depending on when the snapshot is viewed, this can produce slight offsets like those seen in the left panel of Fig.~\ref{fig: StreamAngularMomentumEvo}.

In the time-dependent tilting disk potential considered here, $L_z$ is no longer conserved. Instead, for rotation about the $x-$axis with frequency $\Omega_{\rm Disk}$, $L_z$---now defined as the component of $\mathbf{L}$ along the disk's tilting normal axis ($z^\prime$ in Fig.~\ref{fig: coord_sys})---evolves as:
\begin{equation}\label{eq: dot_p_phi}
\dot{L}_z = \Omega_{\rm Disk}\left[z\frac{L_z}{R}\sin\phi - \left(zp_R - Rp_z \right)\cos\phi\right],
\end{equation}
where $p_R$ is the radial component of the cartesian momentum in cylindrical coordinates (co-rotating with the disk) and $p_z$ is the vertical component (see Appendix~\ref{app: EqnsofMotion} for derivation). 

From Eq.~\ref{eq: dot_p_phi}, it is immediately clear that trajectories are no longer confined to the meridional plane in the tilting disk scenario. With an additional degree of freedom in $L_z$, a tilting disk can produce orbits with a wider range of inclinations than possible in a static axisymmetric potential. Indeed, this is the behavior we find in the right panel of Fig.~\ref{fig: StreamAngularMomentumEvo}, where we take the same final progenitor locations in the left panel and generate streams in the tilting disk potential with $\mathbf{\Omega}_{\rm Disk} = (15~\rm{deg}/\rm{Gyr})\hat{\mathbf{x}}$. Geometrically, streams produced in the time-dependent potential can no longer be characterized as a $\psi \approx \rm{constant}$ cone. Rather, the orbital poles of each stream are tilted, spanning a wider range in inclination angles.

Streams with a small $v_R/v_z$ ratio are more tilted compared to those generated with a higher $v_R/v_z$ ratio.  The former case corresponds to streams on more inclined orbits, while the latter case corresponds to streams on orbits that are more closely aligned with the disk. This can be understood from Eq.~\ref{eq: dot_p_phi}: orbits with large vertical oscillations (i.e., large $z_{\rm max}$ at $p_z = 0$, and large $p_{z,{\rm max}}$ at $z = 0$) will experience large values of $\dot{L}_z$. Thus, streams on more inclined orbits with large vertical excursions should show signatures of vertical torques in their pole vector distribution, provided that the disk has experienced significant angular momentum evolution over the stream's formation. In Appendix~\ref{sec: small_vertical_osc}, we show that orbits with small vertical excursions roughly conserve $L_z$ in a corotating frame, which explains why the teal stream in Fig.~\ref{fig: StreamAngularMomentumEvo} appears very similar to its static potential counterpart. As a result, streams on moderately inclined orbits are more sensitive probes of stellar disk tilting compared to those on less inclined orbits.

The right panel of Fig.~\ref{fig: StreamAngularMomentumEvo} also shows the orbital pole vectors of each stream progenitor in the tilting disk potential (gray curves), as viewed in a corotating frame. As before, purple symbols indicate the normal vectors to a series of planes fit to the spatial location of $20~\rm{deg}$ stream segments in position space. Interestingly, the alignment between the stream orbital poles, its spatial planes, and the progenitor's orbital poles are all maintained for the tilting rates considered in this work. While a 15$~\rm{deg}/\rm{Gyr}$ tilt represents a substantial change in the disk's orientation over a stream's lifetime, this tilting rate corresponds to a 24~\rm{Gyr} period, which is still much longer than an orbital timescale. In this regime, the stream has time to adjust gradually to the new position of the disk, maintaining alignment between its spatial plane and orbital pole vectors. We also see this behavior in the stream-track and proper-motion alignment in Fig.~\ref{fig: HeatingCoolingStream}. Thus, the angular momentum evolution of the disk does not necessarily contribute to a stream track-velocity misalignment. Observationally this is interesting, since even streams without a track-velocity misalignment may be actively affected by the disk's tilt.

In summary, the orbital pole distribution of inclined, eccentric streams provides a sensitive probe of the disk's angular momentum evolution. While a static axisymmetric potential produces streams that are  well-localized to a narrow band in $\psi$, introducing time-dependence in the disk can lead to vertical torques which should systematically tilt the pole vectors of nearby streams with moderately inclined orbits. Importantly, the effect of the disk on a population of streams is global, so streams at different locations and inclinations should experience a consistent tilt in their orbital pole distribution. While a measurement of orbital poles requires 6D phase-space information, Fig.~\ref{fig: StreamAngularMomentumEvo} indicates that the spatial planes occupied by each stream coincides with their orbital pole vectors. Thus, 6D phase-space data is not strictly required to infer the disk's angular momentum evolution and 3D spatial data of stream stars alone can be used to investigate whether the data are consistent with a tilting disk. We leave an analysis of real kinematic and 3D spatial data to future work, though note that this information is available for several Milky Way streams (e.g., \citealt{2023arXiv231116960S}).

\section{Dynamical Effect on Orbits and Stream Width}\label{sec: effect_stream_width}
In this section, we discuss the dynamics giving rise to the fanning and narrowing of stellar streams observed in our simulations and illustrated in Figs.~\ref{fig: HeatingCoolingStream}-\ref{fig: VaryOmegaDisk_summary}. The narrowing effect is not easily reproduced with other time-dependent perturbations like the stellar bar, which tends to produce stream-fanning instead \citep{2016MNRAS.460..497H,2017NatAs...1..633P,2020ApJ...889...70B}. We therefore focus on stream narrowing as a unique observational prediction of the disk's angular momentum evolution. To characterize the scenarios under which a stream will fan out or become more narrow, we also provide a simple description of how orbits are modified by a tilting disk.

\begin{figure}
    \centering\includegraphics[scale=1.75]{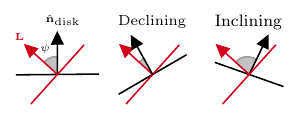}\caption{Schematic illustration of how a tilting disk changes the inclination of an orbit. The disk (black line) is viewed edge-on, and the angular momentum vector of an orbit is shown as the red arrow. The normal to the disk is $\hat{n}_{\rm disk}$. If the disk's tilt aligns $\hat{n}_{\rm disk}$ with $\mathbf{L}$ (middle), the orbital inclination is reduced. We refer to this as the declining effect. If the disk's tilt misaligns $\hat{n}_{\rm disk}$ with $\mathbf{L}$ (right), the orbital inclination increases. We refer to this as the inclining effect. Orbits that have undergone the declining or inclining effects experience different torques from the disk throughout their evolution.  }
    \label{fig: tilting_graphic}
\end{figure} 

\begin{figure*}
    \centering\includegraphics[scale=0.52]{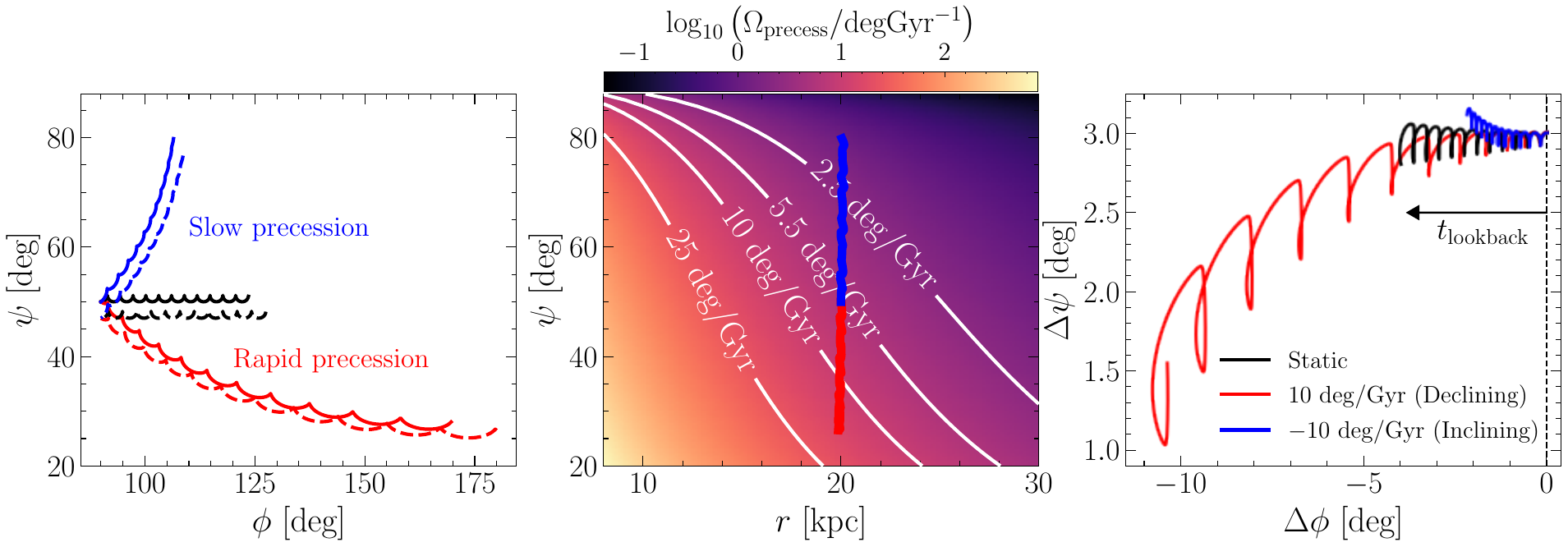}
    \caption{The polar and azimuthal evolution of nearby particles in a static potential and tilting disk potential. \textbf{Left Panel:} Black curves represent the evolution of orbital poles for two nearby particles in the fiducial static potential separated by an initial inclination angle of $3~\rm{deg}$. Blue and red curves show a pair of orbits in the tilting disk potential subject to the declining effect (red) and the inclining effect (blue). The total amount of precession ($\phi$ evolution) across the three cases is distinct, with the inclined orbit precessing less than the declined orbit over the same timescale. \textbf{Middle Panel:} The heatmap shows the precession frequency (in $\log_{10}$) for a circular orbit as a function of galactocentric distance ($r$) and inclination ($\psi$) in the fiducial static potential. Precession frequencies are comparable to the expected disk-tilting rates from cosmological simulations. Red and blue lines represent the evolution of the declined and inclined orbits in this space. \textbf{Right Panel:} The difference in orbital poles between the orbital pairs shown in the left panel. The differential precession and nutation frequencies differ significantly between the three orbital pairs. }
    \label{fig: RadializeCircularizeDemo}
\end{figure*} 

\subsection{Evolution of Orbital Inclination in a Tilting Disk Potential}
In this section, we consider how the angular momentum pole vector of an orbit, and therefore the orbital inclination, evolves in response to a tilting disk. We illustrate the general behavior with a few test-particle orbits in \S\ref{sec: test_particle} and orbit ensembles (i.e., streams) in \S\ref{sec: orbit_ensembles}.

\subsubsection{Test Particle Orbits}\label{sec: test_particle}
In a static axisymmetric potential, the inclination of a stream progenitor determines the differential precession and nutation frequencies experienced by its tidal tails. The characteristic width of a stream is therefore related to the progenitor's orbital inclination \citep{2016MNRAS.461.1590E}.

In the context of a tilting disk, the same relations are complicated by the presence of pseudo-torques (e.g., Eq.~\ref{eq: dot_p_phi}), which act to modify the nutation and precession frequencies in a corotating frame. We can determine the typical effect on orbits in the tilting disk potential by considering the simplified picture in Fig.~\ref{fig: tilting_graphic}. In this illustration, the disk (black horizontal line) is viewed edge on with normal vector given by $\hat{\mathbf{n}}_{\rm disk}$. We will now work in a co-rotating frame, so that the z-axis is aligned with $\hat{\mathbf{n}}_{\rm disk}$ (aligned with $z^\prime$ in Fig.~\ref{fig: coord_sys}). Consider an orbit whose plane is shown in red, with total angular momentum vector $\mathbf{L}$. The angle between the disk and the orbital plane is $\psi$. If we tilt the disk so that its midplane aligns with the orbital plane (i.e., a counter-clockwise rotation in Fig.~\ref{fig: tilting_graphic}), then the orbital inclination ($\psi$) decreases and $L_z = \mathbf{L}\cdot \hat{\mathbf{n}}_{\rm disk}$ will tend to increase in magnitude. We refer to this as the \emph{declining} effect, since the orbital plane becomes more aligned with the disk.

Alternatively, if the disk tilts in the opposite direction (i.e., clockwise), the orbital plane becomes misaligned with the disk's midplane. In this case, the inclination of the orbit increases, and $L_z$ will tend to decrease in magnitude. We refer to this as the \emph{inclining} effect, since the inclination of the orbit relative to the disk increases.

This migration of orbital planes towards less inclined or more inclined orbits is illustrated in Fig.~\ref{fig: RadializeCircularizeDemo}, left panel. In this panel, we plot the time-evolution of orbital pole vectors for a set of orbits. The two black lines (solid and dashed) represent two similar orbits integrated in the fiducial static potential separated by a small inclination angle of $\Delta \psi = 3~\rm{deg}$, chosen to consider how similarly inclined orbits will diverge or converge. The orbits undergo precession ($\phi$ evolution) and nutation ($\psi$ oscillations) due to the axisymmetric disk potential. We then consider how these orbits evolve if we take the same final phase-space coordinates and integrate backwards while aligning the disk with the orbital plane (declining) and misaligning the disk with the orbital plane (inclining). These two cases are shown in red and blue, respectively, in a frame co-rotating with the disk. As anticipated, orbits subject to the inclining effect achieve larger inclinations, while those subject to the declining effect become more aligned with the disk's plane. 

The left panel of Fig.~\ref{fig: RadializeCircularizeDemo} also demonstrates that the inclining and declining of orbits has opposite effects on the angular momentum precession frequency. In particular, the inclined orbits precess by only a small amount, $\Delta \phi \approx 5~\rm{deg}$ over 4~Gyr, while the declined orbit precesses by a large amount, $\Delta \phi \approx 80~\rm{deg}$, over the same period. This effect is due to the sensitivity of the precession frequency to the orbital inclination. 

The middle panel of Fig.~\ref{fig: RadializeCircularizeDemo} shows the angular momentum precession frequency for orbits in the fiducial static potential, numerically estimated as a function of galactocentric distance $r$ (x-axis) and orbital inclination $\psi$ (y-axis) from a grid of approximately circular orbits. The white curves indicate contours of constant precession frequency. We find that increasing the orbital inclination tends to decrease the orbital precession frequency, in agreement with \citet{2016MNRAS.461.1590E}.  

\begin{figure*}
    \centering
    \includegraphics[scale=0.65]{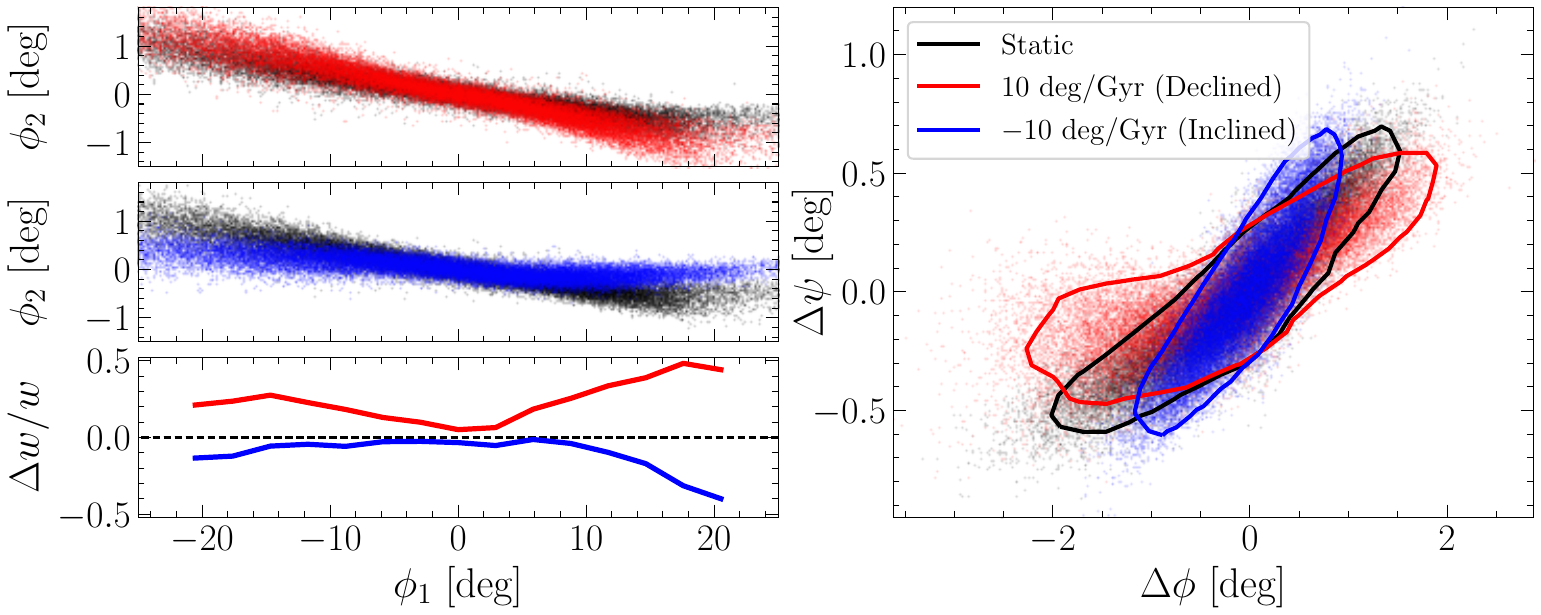}
    \caption{The stream equivalent of Fig.~\ref{fig: RadializeCircularizeDemo}. Streams sharing the same final progenitor location, subject to the declining effect (red) and the inclining effect (blue). The black points correspond to the static potential stream. The declining stream fans out, as indicated by its enhanced width in the bottom left panel. The inclined stream is more narrow as a result of the tilting disk. The orbital pole distribution for the three streams is shown in the right panel. This stream experiences a smaller amount of fanning and narrowing compared to the stream in Fig.~\ref{fig: HeatingCoolingStream} due to the lower tilting rate considered, and the stream's larger distance of $20~\rm{kpc}$. Still, the orbital pole distribution of the inclining stream is less extended azimuthally compared to the declining stream, in-line with expectations from Fig.~\ref{fig: RadializeCircularizeDemo}. }
    \label{fig: MoreCircularStream}
\end{figure*}

Also in the middle panel of Fig.~\ref{fig: RadializeCircularizeDemo}, we overplot the solid blue (inclining) and solid red (declining) orbits from the left panel. As anticipated from the left panel, the inclining orbit moves towards regions of the potential where the precession frequency is lower, while the declining orbit migrates to regions with more rapid precession. Additionally, the timescale of precession is similar to the timescale of the disk's expected angular momentum evolution, since a roughly circular orbit at $r \approx 20~\rm{kpc}$ with $\psi \approx 35~\rm{deg}$ will precess with a frequency of $10~\rm{deg}/\rm{Gyr}$. The similarity of timescales permits the orbital inclination of a star today to differ significantly from its inclination in the past as a result of the disk's angular momentum evolution. While this result is not specific to streams, it does have a direct impact on stream width as we will discuss in \S\ref{sec: precession_nutation_freqs}. 

So far we have considered single particle orbits. Stellar streams are, however, an ensemble of nearby orbits. To determine how orbital inclining and declining influence the ensemble properties of nearby orbits, we plot the difference in orbital pole angles $\Delta\phi$ and $\Delta\psi$, over time, in the right panel of Fig.~\ref{fig: RadializeCircularizeDemo} for each orbital pair shown in the left panel. From this exercise, we see that in the static potential (black), the orbital poles diverge in azimuth and exhibit small oscillations in their differential inclination angle. Compared to the orbits in a static potential, the inclining orbital pair experiences significantly less differential precession  while the declining orbital pair experiences significantly more differential precession. There are also differences in the differential inclination angle of the orbits, with the inclining pair becoming slightly more separated in inclination ($\Delta \psi$ grows), while the declining pair becomes significantly more aligned in inclination ($\Delta \psi$ shrinks).

We might anticipate, then, that streams subjected to the inclining effect will tend to appear narrow in their azimuthal pole vector distribution, while those that are subjected to the declining effect will appear more extended. Conversely, the dispersion in inclination angles for a declining stream will
tend to be less compared to the inclination dispersion for the inclined stream, since the declining orbital pair in Fig.~\ref{fig: RadializeCircularizeDemo} experiences less differential nutation.

\subsubsection{Orbit Ensembles: Streams}\label{sec: orbit_ensembles}
To demonstrate how the behavior of the orbital pairs in Fig.~\ref{fig: RadializeCircularizeDemo} generalizes to streams, we generate a new set of streams on roughly circular orbits ($75\%$ of the local circular velocity at apocenter) with final progenitor orbital pole vectors similar to the experiment in Fig.~\ref{fig: RadializeCircularizeDemo} ($\psi \sim 50~\rm{deg}$ and $\phi \sim 90~\rm{deg}$). We tilt the disk around $\pm {\mathbf{x}}$ with $\Omega_{\rm Disk} = 10~\rm{deg}/\rm{Gyr}$, and the distance to the progenitor from the Galactic center is $\sim 20~\rm{kpc}$. The orbital pole distribution for this set of streams is shown in the right panel of Fig.~\ref{fig: MoreCircularStream}, where the $+{\mathbf{x}}$ rotation leads to lower past inclinations (i.e., declining), and the $-{\mathbf{x}}$ rotation leads to larger past inclinations (i.e., inclining).

As expected from our experiment in Fig.~\ref{fig: RadializeCircularizeDemo}, the inclined stream (blue) is suppressed in its $\Delta \phi$ spread compared to the static potential stream (black), while the declined stream is more extended in $\Delta\phi$. Differences in the inclination distribution of each stream's pole vectors are subdominant to differences in azimuth.

 Differences in the total amount of differential precession and nutation are reflected in the width of streams, since a large amount of differential precession or nutation will scatter the orbital planes of individual stars in a stream. This is demonstrated in the left panels of Fig.~\ref{fig: MoreCircularStream}, where the static potential (black), declining (top), and inclining (middle) streams are shown in the same $\phi_1-\phi_2$ frame. The fractional change in stream width (relative to the unperturbed static model) is shown in the bottom panel. The inclining stream is indeed thinner than the static potential stream, while the declining stream is more diffuse. This can be understood by the orbital pole distribution of the streams, since the angular momentum ``footprint" of the inclining stream has a small extent in $\Delta \phi$ compared to the static model, while the declining stream is comparatively more extended  in its orbital pole distribution. 

 In this section, we have demonstrated that disk tilting can produce narrow streams due to the inclining effect and more diffuse streams due to the declining effect. The former effect tends to align orbital poles, while the latter leads to the divergence of orbital poles. In the following section (\S\ref{sec: dynamics_stream_width}), we discuss the dynamical origin of this behavior.

\subsection{Modeling the Dynamics of Stream Width in a Tilting Disk Potential}\label{sec: dynamics_stream_width}
The declining and inclining of orbits has a direct impact on the rate of precession and nutation of the orbital angular momentum vector.
We discuss how both scenarios give rise to stream fanning and narrowing below. 
We first provide a general description of the angular momentum frequency evolution in response to the tilting disk and then specialize to the case of precession and nutation in \S\ref{sec: precession_nutation_freqs}.

Consider a particle stripped from the progenitor cluster. The orbital inclination of the progenitor at stripping time is $\psi_{\rm prog,0}$, and the inclination of the stripped particle is $\psi_{\rm strip,0}$. The initial angular momentum frequencies (i.e., precession or nutation) are $\Omega_{\rm prog,0}$ and $\Omega_{\rm strip,0}$, respectively.

We now determine how the frequency difference, $\Delta \Omega = \Omega_{\rm strip} - \Omega_{\rm prog}$, between the two particles grows or shrinks as a result of a tilting disk potential. We examine the short-term evolution  of $\Delta \Omega$ as a function of lookback time rather than future time, since a consistently smaller $\Delta \Omega$ \emph{in the past} will produce a more narrow stream \emph{today}. Parameterizing as a function of lookback time also allows us to more readily draw connections to the data. For clarity, we will assume that $\psi_{\rm prog,0}, \psi_{\rm strip, 0} \in [0,90)~\rm{deg}$, though this is not strictly necessary. 

The magnitude of the tilting disk's effect on an orbit depends on the relative orientation of the tilting axis to the orbit's angular momentum vector. For simplicity, we will now consider a scenario where $\mathbf{L}$ is orthogonal to $\hat{\mathbf{n}}_{\rm tilt}$. A description for an arbitrary tilting axis and its effect on orbits is provided in Appendix~\ref{app: arb_tilting_axis}. With our assumption of an orthogonal angular momentum vector and tilting axis, if the disk tilts by some increment $\Delta \psi$ over a lookback time interval $\Delta t_{\rm back} > 0$, a particle with the initial precession or nutation frequency $\Omega_0$ and orbital inclination $\psi_0$ will acquire the new frequency

\begin{equation}\label{eq: omega_expression}
    \Omega_{\rm new} \approx \Omega_0 + \frac{d\Omega}{d\psi}\Big\vert_{\psi_0} \frac{d\psi}{dt_{\rm back}}\Delta t_{\rm back}.
\end{equation}
where $\frac{d\psi}{dt_{\rm back}} \equiv \Omega_{\rm Disk}$ for the geometry considered.
Importantly, $d\psi/dt_{\rm back} > 0$ if the orbital inclination increases (i.e., inclining) with increasing lookback time, and $d\psi/dt_{\rm back} < 0$ if the orbital inclination decreases (i.e., declining) with increasing lookback time. From Eq.~\ref{eq: omega_expression}, the frequency difference between the stripped particle and progenitor is
\begin{equation}\label{eq: freq_diff_precession}
\begin{split}
    \Delta \Omega_{\rm new} &\approx \left(\Omega_{\rm strip,0} - \Omega_{\rm prog, 0 } \right)\\ &+ \left[\Omega^\prime\left(\psi_{\rm strip, 0}\right) - \Omega^\prime\left(\psi_{\rm prog, 0}\right) \right] \frac{d\psi}{dt_{\rm back}} \Delta t_{\rm back} \\
    &= \Delta\Omega + \Delta \Omega^\prime \frac{d\psi}{dt_{\rm back}} \Delta t_{\rm back} \\
    &= \Delta \Omega\left[1 + \frac{\Delta \Omega^\prime}{\Delta \Omega} \frac{d\psi}{dt_{\rm back}}\Delta t_{\rm back} \right] \\
    &= \Delta \Omega \left[1 \pm \Big\vert \frac{d}{d\psi}\ln\left(\vert \Omega^\prime  \vert \right)\Big\vert\frac{d\psi}{dt_{\rm back}}\Delta t_{\rm back} \right],
    \end{split}
\end{equation}
where $\Delta \Omega$ is the frequency difference in the static potential, and $\Delta \Omega^\prime$ is the derivative of the frequency difference with respect to $\psi$. In the final line of Eq.~\ref{eq: freq_diff_precession}, we have used $f^{(k)}(x) \Delta x \approx \Delta f^{(k-1)}(x)$, where $f^{(k)}(x)$ is the $k^{\rm th}$ derivative of $f$ with respect to $x$ (i.e., $\Delta\Omega^\prime\approx \Omega^{\prime\prime}(\psi) \Delta \psi$). For a time-dependent frequency, the angular difference between orbital poles is approximately
\begin{equation}\label{eq: angle_integral}
    \Delta \theta(t) \approx \int\limits_t^{t_{\rm strip}} dt_{\rm back}^\prime \Delta\Omega_{\rm new}\left(t_{\rm back}^\prime\right),
\end{equation}
where $t$ and $t_{\rm strip}$ are lookback times and $t_{\rm strip} > t$. We have assumed that $\Delta\theta(t_{\rm strip})$ is negligible. In the following section, we consider how Eqs.~\ref{eq: freq_diff_precession} and \ref{eq: angle_integral} behave for the precession and nutation frequencies of inclining versus declining stream progenitors. 

\subsubsection{Precession and Nutation}\label{sec: precession_nutation_freqs}
\begin{figure}
    \centering\includegraphics[scale=.5]{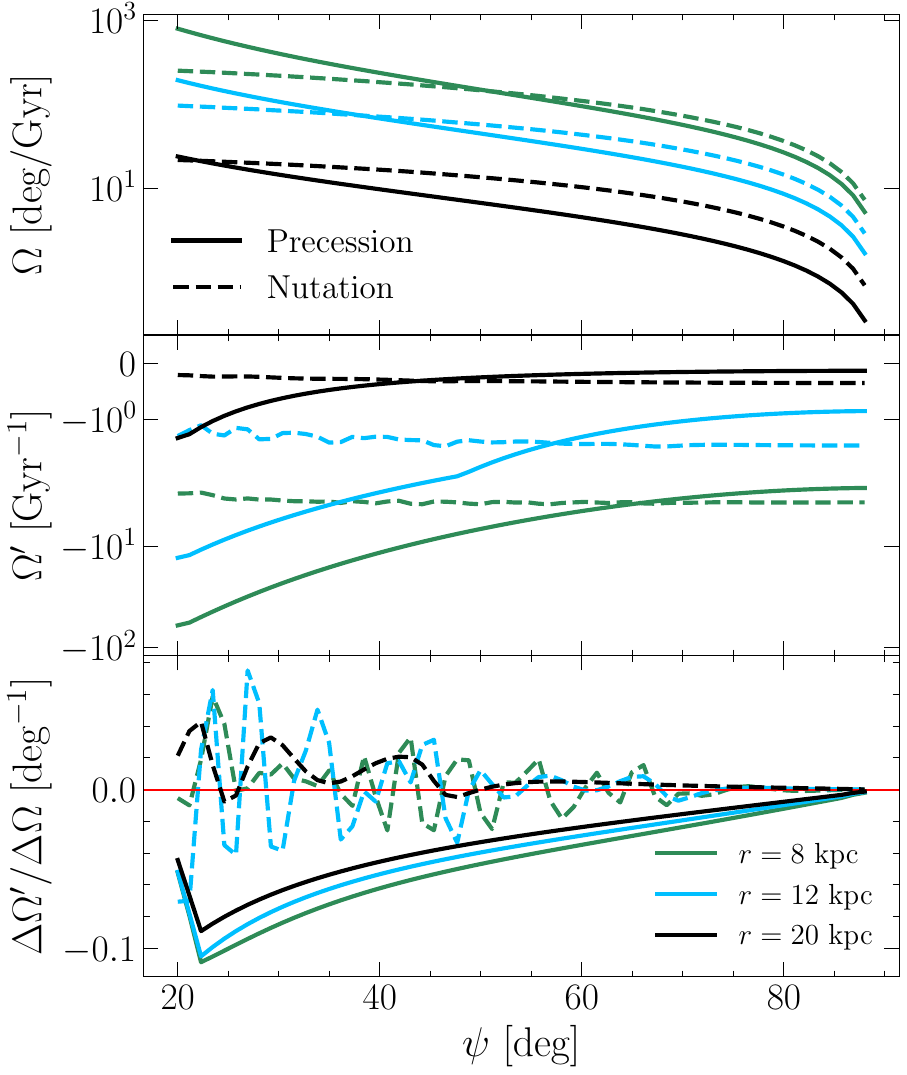}
    \caption{The relevant quantities needed to evaluate Eq.~\ref{eq: freq_diff_precession}. From a series of circular orbits, in the top panel we estimate the precession (solid curves) and nutation (dashed curves) frequencies as a function of orbital inclination, $\psi$, and distance, $r$. The middle panel represents numerically estimated derivatives of the top curves, where $\Omega^\prime \equiv d\Omega/d\psi$. In the bottom panel, $\Delta$ represents a difference between neighboring frequencies (or frequency derivatives) separated by small differences in their inclination. The ratio $\Delta\Omega^\prime / \Delta \Omega$ controls the rate of change of the differential precession and nutation frequencies when nearby particles are perturbed from their initial inclination.}
    \label{fig: Omega_derivs}
\end{figure}

We now discuss how Eq.~\ref{eq: freq_diff_precession} modifies the precession and nutation frequencies of the orbital angular momentum vector. In particular, we would like to consider whether inclining or declining orbits tend to increase the value of $\Delta \Omega$, or suppress $\Delta \Omega$. We can determine a qualitative picture by considering our fiducial static potential model consisting of a Miyamoto-Nagai disk and spherical NFW halo. For this pairing, the orbital precession and nutation frequencies and their derivatives are shown in Fig.~\ref{fig: Omega_derivs}. To generate this figure, we integrate orbits at a given inclination ($\psi$) and 
radius, 
with initial velocity equal to the local circular velocity. To estimate precession frequencies for each orbit, we fit the slope of the precession angle, $\phi(t)$, of the orbital angular momentum against time. The nutation frequency is itself oscillatory, so to obtain a rough estimate of $\Omega_{\rm nutate}$ for each orbit, we fit a sinusoid to the nutation angle, $\psi(t)$, against time, and take the maximum slope of the sinusoid as $\Omega_{\rm nutate}$. The derivatives of these curves (middle panel) are estimated numerically. 

The top panel of Fig.~\ref{fig: Omega_derivs} shows that the precession (solid curves) and nutation (dashed curves) frequencies are unsurprisingly larger for closer in orbits. This is due to the larger torque experienced by a test particle orbiting near the disk. The middle panel of Fig.~\ref{fig: Omega_derivs} shows the numerically estimated derivatives $\Omega^\prime \equiv d\Omega/d\psi$. This quantity tells us the sensitivity of the precession and nutation frequencies to small changes in inclination and enters our linear perturbation analysis through Eq.~\ref{eq: omega_expression}. Because orbits with closer apocenters have a larger magnitude of $\Omega^\prime$ for both precession and nutation, we expect that nearby streams (i.e., those with closer apocenters) should experience the largest change in their on-sky tracks, since the $\Omega^\prime$ profile gives us the first order effect of an orbit to a perturbation in $\psi$.

The bottom panel of Fig.~\ref{fig: Omega_derivs} can be used to understand the behavior of nearby orbits and therefore stream width. In this panel, the ratio $\Delta\Omega^\prime/\Delta \Omega$ (which is one of the main prefactors in Eq.~\ref{eq: freq_diff_precession}) is estimated from the top two panels. From this panel, we can see that $\Delta\Omega^\prime \ \rm{and} \ \Delta\Omega$ consistently take on opposite signs for precession, so that their ratio is negative. Therefore, the correct sign for the second term in the brackets of Eq.~\ref{eq: freq_diff_precession} is negative for precession frequencies. For the nutation frequency, the behavior is highly oscillatory around zero, especially for more circular orbits with closer apocenters (i.e., $r\lesssim 20~\rm{kpc}$). We therefore refrain from choosing a single sign for the bracketed term for the case of nutation. However, for more distant circular orbits, we expect the correct sign choice to be positive (since $\Delta\Omega^\prime/\Delta\Omega$ is typically positive for $r \sim 20~\rm{kpc}$ in Fig.~\ref{fig: Omega_derivs}). Specializing to the cases of precession and nutation, we can rewrite Eq.~\ref{eq: freq_diff_precession} as

\begin{small}
\begin{equation}\label{eq: specialized_precession_difference}
\begin{split}
    \Delta\Omega^{\rm precess}_{\rm new} &\approx \Delta\Omega_{\rm precess}\left[1 - \Big\vert \frac{d}{d\psi} \ln\left(\vert \Omega_{\rm precess}^\prime \vert\right) \Big\vert \frac{d\psi}{dt_{\rm back}}\Delta t_{\rm back} \right] \\
    \Delta\Omega^{\rm nutate}_{\rm new} &\approx \Delta\Omega_{\rm nutate}\left[1 \pm \Big\vert \frac{d}{d\psi} \ln\left(\vert \Omega_{\rm nutate}^\prime \vert\right) \Big\vert \frac{d\psi}{dt_{\rm back}}\Delta t_{\rm back} \right].
\end{split}
\end{equation}\end{small}
Therefore, we can predict that a tilting disk which causes the progenitor to increase its inclination in the past ($d\psi/dt_{\rm back} > 0$; inclining) will tend to suppress the differential precession frequency. If the tilting disk reduces the inclination of the progenitor's orbit ($d\psi/dt_{\rm back} < 0$; declining), we expect the differential precession frequency to be larger in the past. On the other hand, we do not expect a consistent effect for the differential nutation frequnecy. 

\begin{figure}
    \centering\includegraphics[scale=.5]{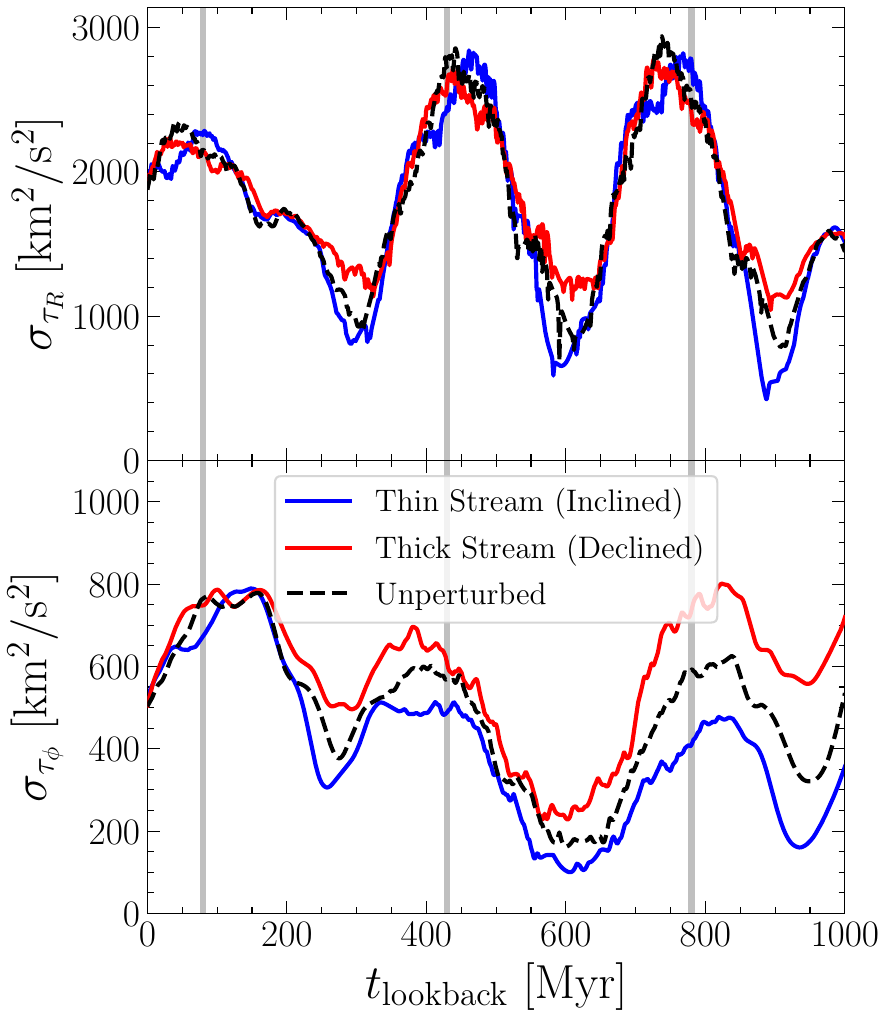}
    \caption{The dispersion in torques as a function of lookback time estimated from $\sim 150$ tracer particles for the stream shown in Fig.~\ref{fig: RadializeCircularizeDemo}. Gray lines represent pericentric passages of the static potential progenitor. Blue and red profiles are for a disk-tilting rate of 10~deg/Gyr. At each timestep, torques are decomposed into a cylindrical radial component ($\tau_R$; responsible for nutation, top panel) and azimuthal component ($\tau_\phi$; responsible for precession, bottom panel), and the standard deviation of the torque distribution is calculated over the ensemble. The inclined stream experiences systematically less azimuthal torque dispersion (bottom panel) than both the unperturbed and declined streams. The relative radial torque dispersion between the three cases is more complicated.  }
    \label{fig: Differential_Torque}
\end{figure} 

The consistency of signs in Eq.~\ref{eq: specialized_precession_difference} for precession has the ability to produce more narrow streams and more diffuse streams, depending on the sign of $d\psi/dt_{\rm back}$ (as seen in Fig.~\ref{fig: MoreCircularStream}). By aligning the orbital poles of an ensemble of nearby orbits, the angular separation between the particles will typically be smaller. By misaligning orbital poles, the angular separation will typically be larger. 

The bottom panel of Fig.~\ref{fig: Omega_derivs} also suggests that more distant streams should experience a lower rate of change in their differential precession frequency, because the ratio $\Delta \Omega^\prime / \Delta\Omega$ is smaller for more distant orbits. Therefore, while distant streams can indeed fan out or narrow in response to a tilting disk, if we assume that the effect of precession is dominant, the magnitude of the change in stream width should be unsurprisingly smaller for more distant streams. This effect is demonstrated with simulations of the streams Pal 5 and Pal 13 in \S\ref{sec: data}. The same is not necessarily true for nutation, with a $\Delta\Omega^\prime/\Delta\Omega$ ratio that oscillates significantly both as a function of inclination and distance in Fig.~\ref{fig: Omega_derivs}. However, because the ratio is typically larger in magnitude for precession, we expect the above argument on stream width to hold true (especially for more circular orbits).

The above discussion can also be viewed in terms of the differential torque experienced by a set of nearby particles. We now demonstrate the relation to differential torque using the streams presented in Fig.~\ref{fig: MoreCircularStream}. We integrate the orbits of $150$ typical particles in the leading and trailing arms of each stream, and compute the azimuthal torque ($\tau_\phi$; responsible for precession) and radial torque ($\tau_R$; responsible for nutation) experienced by each particle, as viewed in a frame co-rotating with the tilting disk. Importantly, the streams considered here are initialized on orbits with $70\%$ of the local circular velocity with an apocenter of around $20~\rm{kpc}$ and pericenter of around $14~\rm{kpc}$. Thus, we expect the oscillations in the bottom panel of Fig.~\ref{fig: Omega_derivs} for nutation to be relevant. A large dispersion in torques (labeled $\sigma_{\tau_R}$ and $\sigma_{\tau_\phi}$) experienced across the ensemble is responsible for producing more diffuse streams. As a function of lookback time, the torque dispersion is illustrated in Fig.~\ref{fig: Differential_Torque} for both $\tau_R$ and $\tau_\phi$. Gray vertical lines represent subsequent pericentric passages of the unperturbed progenitor when the total torque is typically largest. 

In the top panel of Fig.~\ref{fig: Differential_Torque}, we find that the differential radial torque is neither consistently larger for the declining nor inclining ensemble. Instead, the two achieve similar maximum values at different times. The inclining progenitor achieves a slightly more distant apocenter than both the unperturbed and declining progenitors, which explains why it has the smallest radial torque dispersion between subsequent pericenters. Similarly, the declining progenitor has slightly closer-in apocenters, and therefore experiences a larger torque dispersion at apocenter. Thus, while there are patterns in the radial torque dispersion at subsequent apocenters, differences in $\sigma_{\tau_R}$ for declining or inclining orbits are not sustained for more than $\sim 100~\rm{Myr}$ outside of each stream's apocenter.

The behavior of the azimuthal torque dispersion in Fig.~\ref{fig: Differential_Torque} is much simpler (bottom panel). In particular, the inclining ensemble experiences systematically less azimuthal torque dispersion compared to the unperturbed ensemble. Meanwhile, the declining ensemble experiences consistently more azimuthal torque dispersion compared to the unperturbed ensemble. This is exactly according to our expectations from Eq.~\ref{eq: specialized_precession_difference}. For only a $10~\rm{deg}/\rm{Gyr}$ tilting rate, the fractional change in the torque dispersion is at the level of $\sim 30\%$ or higher. This indicates that even moderate disk tilting rates can significantly alter the differential torque experienced by an ensemble of nearby orbits, and therefore the morphology of a cluster's tidal tails.

\section{Sensitivity of Milky Way Streams to a Tilting Disk}\label{sec: sensitivity}
We now connect our theoretical predictions for the tilting disk scenario to observations. We first consider the sensitivity of Milky Way globular clusters to stream fanning and narrowing in \S\ref{sec: predict_fan_narrow} and then highlight predictions for the clusters Pal 5 and Pal 13 in \S\ref{sec: data}. In \S\ref{sec: Pal5_Data}, we compare our prediction for Pal 5 to actual measurements of its tidal tails. In \S\ref{sec: implications}, we demonstrate the implications of a tilting disk for reconstructing the gravitational potential. 

\subsection{Predictions for Fanning and Narrowing}\label{sec: predict_fan_narrow}
\begin{figure}
    \centering
    \includegraphics[scale=0.5]{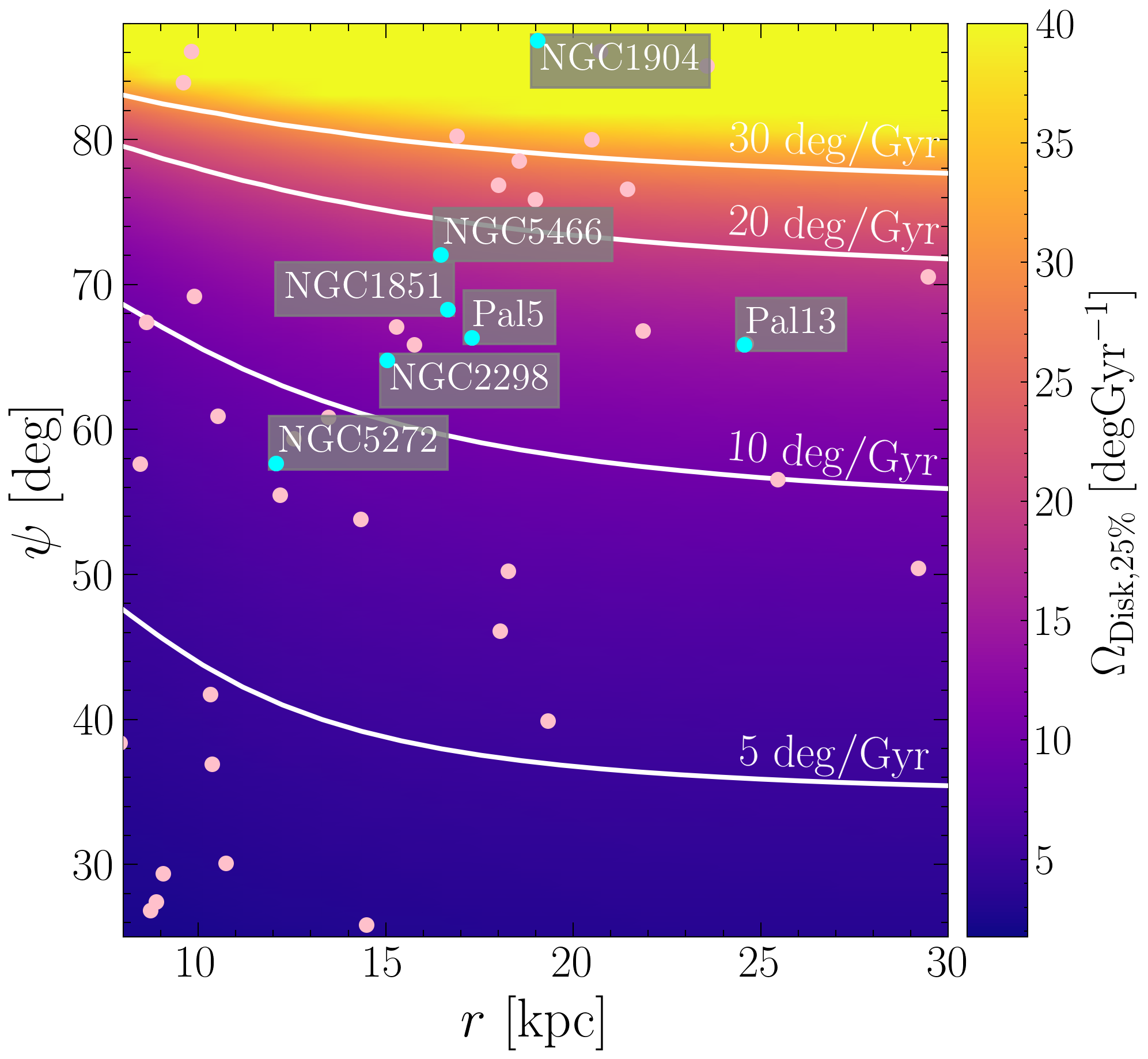}
    \caption{At what tilting rate are streams sensitive to fanning and narrowing? This figure illustrates the tilting rate required to change the differential precession frequency by 25\% per Gyr (derived from Eq.~\ref{eq: 25_percent_change}). Pink points indicate the galactocentric distance and inclination of globular clusters from \citet{2019MNRAS.484.2832V}. Those with known tidal tails are shaded cyan. This figure demonstrates that streams on inclined orbits with apocenters $\sim 20~\rm{kpc}$ should experience appreciable fanning in response to stellar disk tilting at realistic frequencies (i.e., $10$--$20~\rm{deg}/\rm{Gyr}$). }
    \label{fig: 25percent_change}
\end{figure}

In this section, we employ the quantitative description provided in \S\ref{sec: dynamics_stream_width} to predict which streams will be most sensitive to fanning and narrowing due to the disk's angular momentum evolution. To simplify our discussion, we focus only on differential precession, which is dominant for streams with more distant apocenters (Fig.~\ref{fig: Omega_derivs}).

Focusing on precession, we can use Eq.~\ref{eq: specialized_precession_difference} to predict the required disk tilting rate, $d\psi/dt_{\rm back}$, to produce appreciable changes in the differential precession frequency. For instance, over a period of 1~\rm{Gyr}, a 25\% change in $\Delta\Omega_{\rm precess}$ requires a disk tilting frequency of 
\begin{small}
\begin{equation}\label{eq: 25_percent_change}
    \Omega_{\rm Disk, 25\%} = \Big\vert \frac{d\psi}{dt_{\rm back}} \Big\vert \approx  \frac{ 1}{4\vert \frac{d}{d\psi} \ln\left(\vert\Omega_{\rm precess}^\prime\vert\right)\vert } \left( \frac{1}{\rm{Gyr}}\right).
\end{equation}\end{small}

To determine numerical values for $\Omega_{\rm Disk, 25\%}$, we integrate a grid of circular orbits over varying inclination angles and radii. The resulting disk-tilting frequency distribution is shown in Fig.~\ref{fig: 25percent_change}. This figure can be interpreted as the required disk tilting rate to incur a $\sim$25\% change (per Gyr) in the differential precession frequency for each distance and inclination pairing. The white curves indicate contours of constant $\Omega_{\rm Disk,25\%}$, and the pink and cyan points show the locations of galactic globular clusters from \citet{2019MNRAS.484.2832V}. The cyan points indicate globular clusters with known tidal tails and are labeled accordingly.

Fig.~\ref{fig: 25percent_change} indicates that relatively small disk tilting rates are required to produce appreciable changes in the differential precession frequency. Considering that $10-15~\rm{deg}/\rm{Gyr}$ is a typical value quoted in the literature (e.g., \citealt{ 2017MNRAS.469.4095E,2023MNRAS.518.2870D}), streams across a large range of inclinations and radii should be sensitive to the disk's angular momentum evolution. Additionally, we have identified the tilting disk as a mechanism to produce both stream fanning and narrowing. Fig.~\ref{fig: 25percent_change} only provides the typical values required to produce an appreciable change in a stream's angular momentum frequency distribution. Whether this change represents a decrease or increase in the differential frequency experienced by each stream depends on the direction of the disk's tilt, which in turn determines the sign of $d\psi/dt_{\rm back}$ in Eq.~\ref{eq: specialized_precession_difference}. 

\begin{figure*}
    \centering
    \includegraphics[scale=0.5]{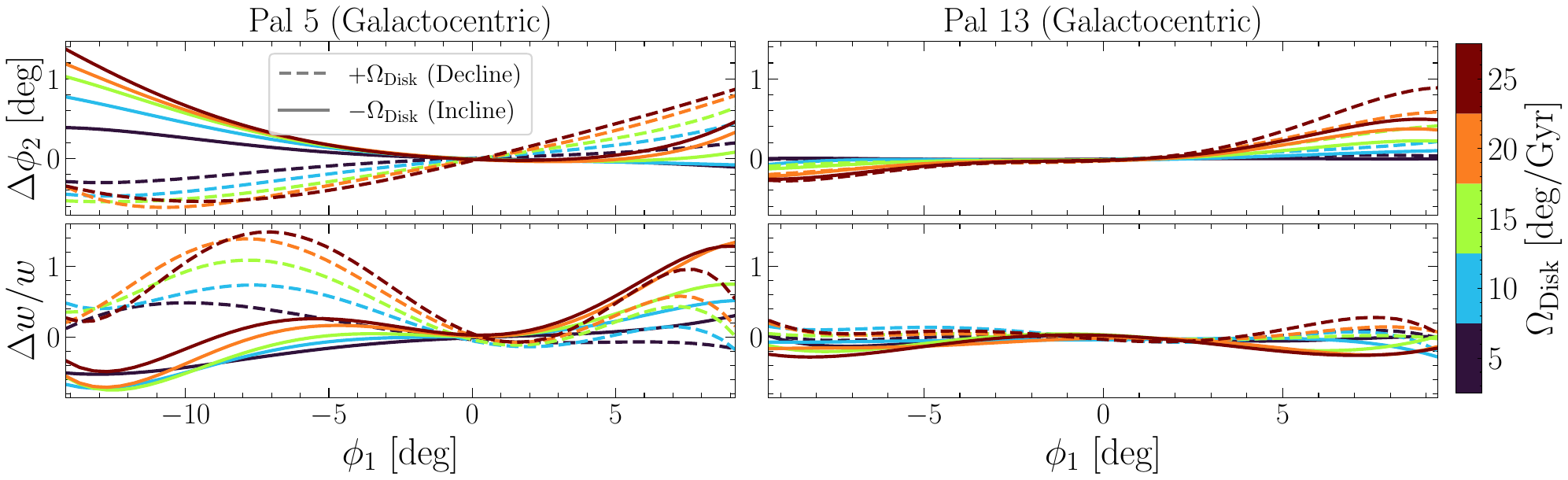}
    \caption{Galactocentric view of Pal 5 and Pal 13's tidal tails, subject to a tilting disk of varying rates. Dashed and solid lines correspond to opposite tilting directions, one producing the declining effect ($+\Omega_{\rm Disk}$) and the other producing the inclining effect ($-\Omega_{\rm Disk}$). Because the apocenteric radius of Pal 13 is much larger than that of Pal 5 ($\sim 50~\rm{kpc}$ versus $\sim 20~\rm{kpc}$), the effect of a tilting disk is more substantial for Pal 5. }
    \label{fig: Pal5Pal13}
\end{figure*} 

For a disk tilting rate of $15~\rm{deg}/\rm{Gyr}$, streams with inclinations near $\psi \sim 70~\rm{deg}$ should be good candidates for fanning and narrowing (i.e., Pal 5 and Pal 13, which we explore in \S\ref{sec: data}). Streams with lower inclinations whose orbits are more aligned with the disk have smaller associated values of $\Omega_{\rm Disk, 25\%} \sim 5~\rm{deg}/\rm{Gyr}$ in Fig.~\ref{fig: 25percent_change}. These streams should all be susceptible to fanning out if the disk's tilt is sustained for several Gyr. This is because for a stream with $\Omega_{\rm Disk, 25\%} \sim 5~\rm{deg}/\rm{Gyr}$, a tilting rate of $15~\rm{deg}/\rm{Gyr}$ integrated over the age of the stream (several Gyr) represents a fractional change in the differential precession frequency of greater than $100\%$ (i.e., the differential precession frequency can decrease to zero, and then increase again). Thus, the response of a population of streams to a tilting disk depends on their location in Fig.~\ref{fig: 25percent_change}. While the width of any given stream might be degenerate with other unknown properties of the potential (i.e., the triaxiality of the halo), Fig.~\ref{fig: 25percent_change} argues that disk tilting produces a predictable effect across a population of streams.

There are several caveats to Fig.~\ref{fig: 25percent_change} and the surrounding discussion. In particular, we have neglected the effect of differential nutation, considered nearly circular orbits, and assumed that $\hat{\mathbf{n}}_{\rm tilt}$ is orthogonal to $\mathbf{L}$. These assumptions can be relaxed, and we provide a discussion of different tilting directions in Appendix~\ref{app: arb_tilting_axis}. However, for more inclined orbits with $\psi\gtrsim 45~\rm{deg}$, Fig.~\ref{fig: Omega_derivs} indicates that the effect of differential nutation should be subdominant to differential precession. Despite our assumptions, we will demonstrate the predictive power of Fig.~\ref{fig: 25percent_change} in determining the sensitivity of streams to fanning and narrowing in \S\ref{sec: data}.

\subsection{The Effect of a Tilting Disk on Pal 5 and  Pal 13}\label{sec: data}
We now consider the effect of a tilting disk on the globular cluster streams Pal 5 and Pal 13. We first illustrate the effect of different tilting rates on the track and width of each stream and then compare our simulations of Pal 5 to actual observations in \S\ref{sec: Pal5_Data}. We focus on Pal 5 and Pal 13 because both clusters have known tidal tails, and the phase-space location of their progenitors have been estimated. Additionally, we choose these clusters based on Fig.~\ref{fig: 25percent_change}, which suggests that the tidal tails of both clusters should be sensitive to fanning and narrowing at similar disk tilting rates  of $\sim 15~\rm{deg}/\rm{Gyr}$. We remind the reader, however, that Fig.~\ref{fig: 25percent_change} does not illustrate the expected magnitude of the narrowing or fanning effect, which we will explore below. 

For the phase-space location of each cluster, we adopt the best-fit values from \citet{2019MNRAS.484.2832V} and assume a total integration time of $4~\rm{Gyr}$ when generating streams. Pal 5 is at a galactocentric distance of $r \approx 18~\rm{kpc}$, while Pal 13 is more distant at $r \approx 25~\rm{kpc}$. In terms of the angles outlined in Fig.~\ref{fig: coord_sys} and neglecting the direction of circulation within the orbital plane (i.e., clockwise versus counterclockwise), Pal 5's angular momentum vector is located at $\psi \approx 66~\rm{deg}$ and $\phi \approx 162~\rm{deg}$. Pal 13's angular momentum vector is similarly inclined, with $\psi \approx 65~\rm{deg}$ and $\phi \approx 45~\rm{deg}$.

 As a fiducial test, we first consider keeping the tilting axis, $\hat{\mathbf{n}}_{\rm disk}$, fixed while varying only the tilting rate, $\Omega_{\rm Disk}$. We tilt the disk around $\pm{\mathbf{y}}$ so that the angular momentum vector of Pal 5 today is nearly orthogonal to $\hat{\mathbf{n}}_{\rm Disk}$. We vary the tilting axis in \S\ref{sec: Pal5_Data}.

Our discussion in \S\ref{sec: effect_stream_width} shows that the location of the orbital angular momentum vector of the progenitor largely determines whether a stream will experience fanning or narrowing in response to a tilting disk. For the case of Pal 5, rotating the disk about $+{\mathbf{y}}$ means that $d\psi/dt_{\rm back} < 0$ initially. This represents the declining effect. Conversely, rotating the disk about $-{\mathbf{y}}$ means that $d\psi/dt_{\rm back} > 0$ initially. This represents the inclining effect. Because Pal 5 has a relatively close-in apocenter ($\approx 18~\rm{kpc}$) and we take the tilting axis to be nearly orthogonal to its angular momentum vector, we expect a tilting disk to have an appreciable effect on the morphology of Pal 5's tidal tails.

Compared to Pal 5, the apocenter of Pal 13 is much more distant ($\approx 50~\rm{kpc}$ in our fiducial static potential, consistent with \citealt{2019MNRAS.484.2832V}) indicating that the effect of differential precession and nutation on the stream should be small (Fig.~\ref{fig: Omega_derivs}, bottom panel). Additionally, the azimuthal location of the cluster's angular momentum vector is at an angle $\alpha \approx 45~\rm{deg}$ from the tilting axis. As discussed in Appendix~\ref{app: arb_tilting_axis}, the effect of the disk's tilt on orbital inclination is largest when $\alpha \approx 90~\rm{deg}$, so smaller values of $\alpha$ typically leave less of an imprint on streams. Taken in combination, the distant apocenter of Pal 13 and the tilting direction considered here should make the stream's width relatively stable to the disk's angular momentum evolution. We do, however, expect a small amount of fanning and narrowing for different rotation directions with $\Omega_{\rm Disk} \approx 15~\rm{deg}/\rm{Gyr}$ from Fig.~\ref{fig: 25percent_change}.

In a galactocentric frame, Fig.~\ref{fig: Pal5Pal13} shows the shift in stream tracks (top row) and the fractional change in stream width (bottom row) for Pal 5 and Pal 13, as a function of different values of $\Omega_{\rm Disk}$. The dashed curves correspond to the $+{\mathbf{y}}$ rotation, while the solid curves correspond to the $-{\mathbf{y}}$ rotation. The effect of the tilting disk is substantially larger for Pal 5 than Pal 13 both for the magnitude of the track offsets and changes in stream width. Changes in the on-sky track of Pal 5 are detectable, with shifts at the level of $0.5$--$1~\rm{deg}$ in $\phi_2$. The effect on the width of Pal 5 is also appreciable, with the $-\Omega_{\rm Disk}$ rotation suppressing the width of the stream by up to $\sim 70\%$ in the trailing arm ($\phi_1 < 0$). For the $+\Omega_{\rm Disk}$ rotation, the stream fans out by up to $\sim 160\%$ in its trailing arm, relative to the unperturbed model. These two rotations give rise to the inclining and declining of the progenitor's orbit, respectively.

For Pal 13, changes in the track of the stream are at the $0.2$--$0.4~\rm{deg}$ level, while changes in the width do not exceed $\approx \pm 20\%$ of the unperturbed model. We do remark, however, that there is a fanning and narrowing effect for Pal 13 as anticipated for different tilting directions. Offsets in the predicted location of the stream track at the level of even $0.1~\rm{deg}$ can be detected in principle, provided that the stream of interest is well characterized. This is not guaranteed, since the membership probability of a possible stream star is typically more uncertain than its angular position on the sky. Still, Fig.~\ref{fig: Pal5Pal13} demonstrates that the amount of variation in a stream's track and width due to the disk's angular momentum evolution can vary significantly depending on the orbital properties of the progenitor.

\begin{figure*}
    \centering
\includegraphics[scale=0.62]{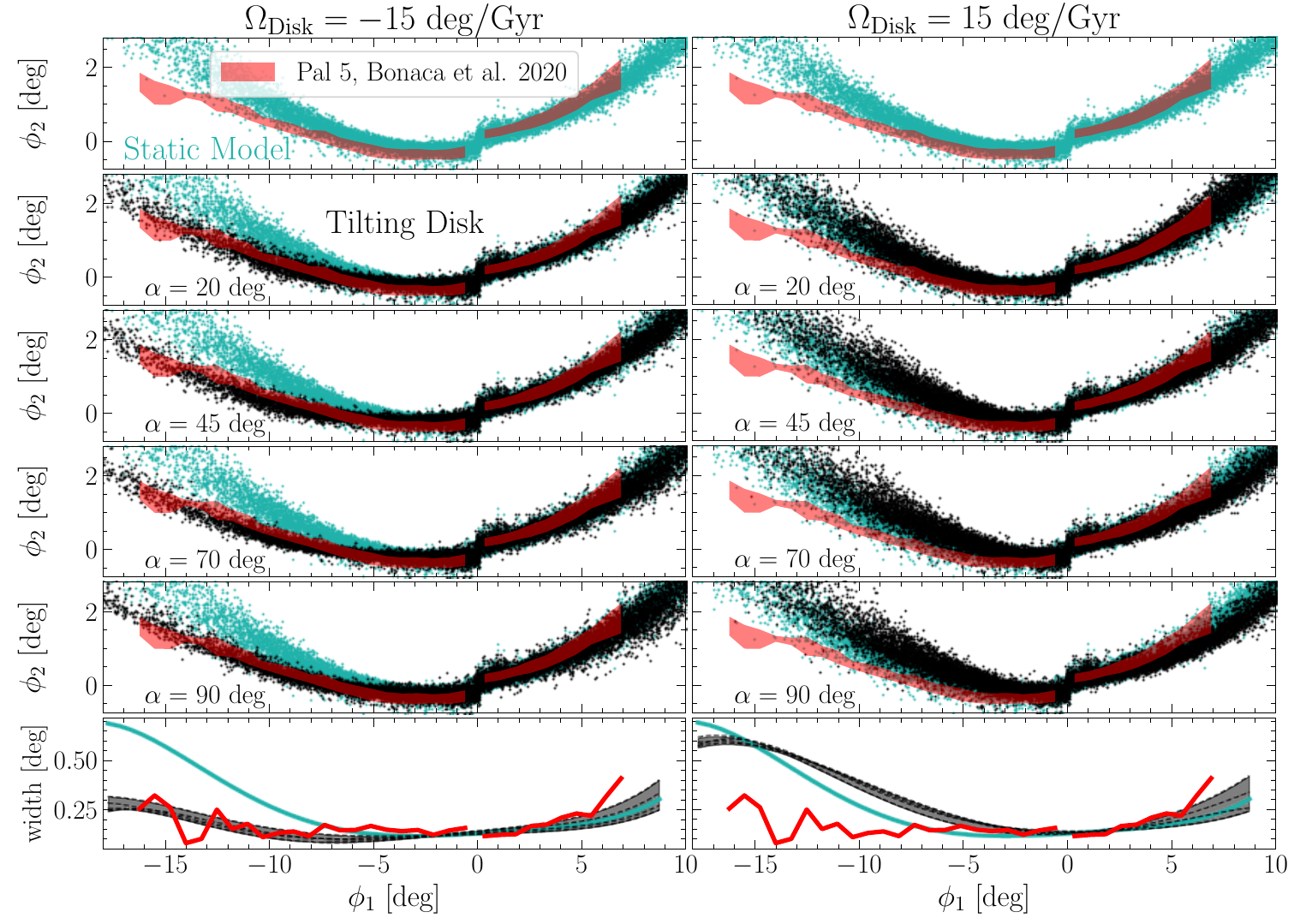}
    \caption{The red band represents data from \citet{2020ApJ...889...70B}, illustrating the track (top 5 rows) and width (bottom row) of Pal 5's tidal tails estimated from DECaLS photometry. Cyan points represent our model for Pal 5 in the fiducial static potential. The static model produces a stream that is too diffuse and misaligned with Pal 5's actual tidal tails. Black points show the effect of our out-of-the-box tilting disk model on Pal 5, as a function of different tilting directions, $\alpha$ (the azimuthal angle between Pal 5's angular momentum vector and the disk's tilting axis). The left column corresponds to a clockwise rotation of the disk about its tilting axis, while the right column represents a counterclockwise rotation. The former causes Pal 5's progenitor to achieve higher inclinations in the past (the inclining effect), while the latter leads to lower orbital inclinations in the past (the declining effect). The inclined streams (left column) provides an excellent match to Pal 5's morphology, both in its on sky track and width. In the right panel, the declined stream is poorly matched to the data, providing a worse description of the stream than the static model.     }
    \label{fig: Pal5}
\end{figure*}

\subsection{Comparison to data: Pal 5}\label{sec: Pal5_Data}
In this section, we compare the observed morphology of Pal 5's tidal tails to mock Pal 5 streams in our fiducial tilting disk simulations. We utilize the $\phi_1-\phi_2$ track and width measurements of Pal 5's tidal tails from \citet{2020ApJ...889...70B}, based on DECaLS grz deep imaging. The track and width characterization is obtained by fitting a series of Gaussians in bins of $\phi_1$ to the CMD-filtered stellar density distribution from the DECaLS data. We adopt a fiducial disk tilting rate of $\Omega_{\rm Disk } = \pm 15~\rm{deg}/\rm{Gyr}$ (where $\pm$ represents a counterclockwise and clockwise rotation, respectively) and consider the effect of varying the azimuthal angle between the tilting axis of the disk ($\alpha$) and angular momentum vector of Pal 5 today. The inclination of the tilting axis is $\beta = 90~\rm{deg}$, representing a rotation around an axis in the plane of the disk. We utilize the fiducial background potential defined in \S\ref{sec: potentials}. The $+15~\rm{deg}/\rm{Gyr}$ rotation tends to align the disk with Pal 5's present-day orbital plane as a function of increasing lookback time, while the $-15~\rm{deg}/\rm{Gyr}$ rotation leads to a misalignment as a function of increasing lookback time.

Our simulation results and comparison to Pal 5's track and width are provided in Fig.~\ref{fig: Pal5}. We view our simulations in the heliocentric $\phi_1-\phi_2$ frame utilized in \citet{2020ApJ...889...70B}, implemented in the \texttt{Gala} package \citep{gala}. The left column corresponds to a clockwise rotation of the disk potential about $\hat{\mathbf{n}}_{\rm tilt}$, while the right column corresponds to a counterclockwise rotation. As discussed in \S\ref{sec: effect_stream_width}, choosing opposite directions for the disk's tilt can produce large differences in the orbit of the progenitor and the resulting stream. The stream in the static disk model is shown by the cyan points, while the black points in each row show the perturbed stream for the respective tilting disk models with the angle $\alpha$ labeled. The red band illustrates the estimated track and width of Pal 5 from \citet{2020ApJ...889...70B}. The bottom row of Fig.~\ref{fig: Pal5} shows the estimated width of Pal 5 in red, the width of the static model in cyan, and the distribution of widths for the tilting disk potentials in gray. 

We first highlight that the simple two-component static potential model consisting of a disk and spherical halo produces a stream that is qualitatively similar to the real Pal 5 stream (top row of Fig.~\ref{fig: Pal5}), but inaccurate under a more detailed inspection. In particular, the $\phi_2$ location of the trailing arm ($\phi_1 < 0~\rm{deg}$) is offset from the data by $\sim 1~\rm{deg}$, and the width of the stream (visualized in the bottom row of Fig.~\ref{fig: Pal5}) is too large for the trailing arm. 

In \S\ref{sec: effect_stream_width}, we discussed how a tilting disk can modify the width of a stream to make it appear more narrow or diffuse. When viewed in a galactocentric frame, we have already shown that Pal 5 can experience significant fanning and narrowing as a result of a tilting disk (Fig.~\ref{fig: Pal5Pal13}). Fig.~\ref{fig: Pal5} provides a heliocentric view of this effect as a function of different tilting directions. Both the stream track and width of our models are similar to the data for the $-15~\rm{deg}/\rm{Gyr}$ tilting disk scenario. Conversely, the $+15~\rm{deg}/\rm{Gyr}$ models provides a worse fit to the actual stream's  track and width compared to the static model. 

The dynamical effect responsible for producing two qualitatively different streams in the left and right column of Fig.~\ref{fig: Pal5} is the inclining and declining of the progenitor's orbit, discussed in \S\ref{sec: effect_stream_width}. In the left column, the rotation direction causes the midplane of the disk to become misaligned with the progenitor's orbital plane in the past (i.e., inclining). This typically leads to a smaller amount of differential precession and therefore more narrow tidal tails. In the right column, the disk becomes increasingly aligned with the orbital plane in the past (i.e., declining),  thereby increasing the rate of differential precession. This tends to produce stream fanning, or a larger stream width.  We find that the former scenario provides a better match to the stream both in its on-sky track and width. Future work could consider other kinematic variations along the stream track (i.e., velocity information), though the dominant difference between the models shown is in the width and track of each stream.

We also find that at least for Pal 5, varying the tilting axis so that it points nearly parallel to ($\alpha = 0~\rm{deg}$) or orthogonal to ($\alpha = 90~\rm{deg}$) the progenitor's angular momentum vector does not produce substantial variations in the morphology and width of the stream. This is because the most significant differences are only found when crossing $\alpha = 180~\rm{deg}$, since this represents the boundary between the inclining and declining effects (illustrated in the left and right columns of Fig.~\ref{fig: Pal5}, respectively). Outside of this interface, the orbital precession frequency of Pal 5 in our fiducial static model is $\sim 10~\rm{deg}/\rm{Gyr}$. Provided that a certain choice of $\hat{\mathbf{n}}_{\rm tilt}$ causes Pal 5 to incline or decline, varying $\alpha$ does not have a large effect since the angular momentum vector already samples a potentially wide range in $\alpha$ values over the stream's lifetime.

Overall, we do not take Fig.~\ref{fig: Pal5} to represent a robust constraint on the angular momentum evolution of the disk. Rather, we claim that at least the morphology of Pal 5 is consistent with the tilting disk scenario, which easily fits the width of the observed stream. While Pal 5 is likely affected by the galactic bar (e.g., \citealt{2017NatAs...1..633P}), the tilting disk scenario provides a novel mechanism to explain the stream's baseline narrow width which is not captured in our static model. Higher-order variations in the width of Pal 5's tidal tails could be explained by the addition of a rotating bar, though we defer an exploration of this effect in combination with a tilting disk to future work. 

\subsection{Implications for Potential Reconstruction using Streams}\label{sec: implications}

\begin{figure*}
    \centering
    \includegraphics[scale=0.55]{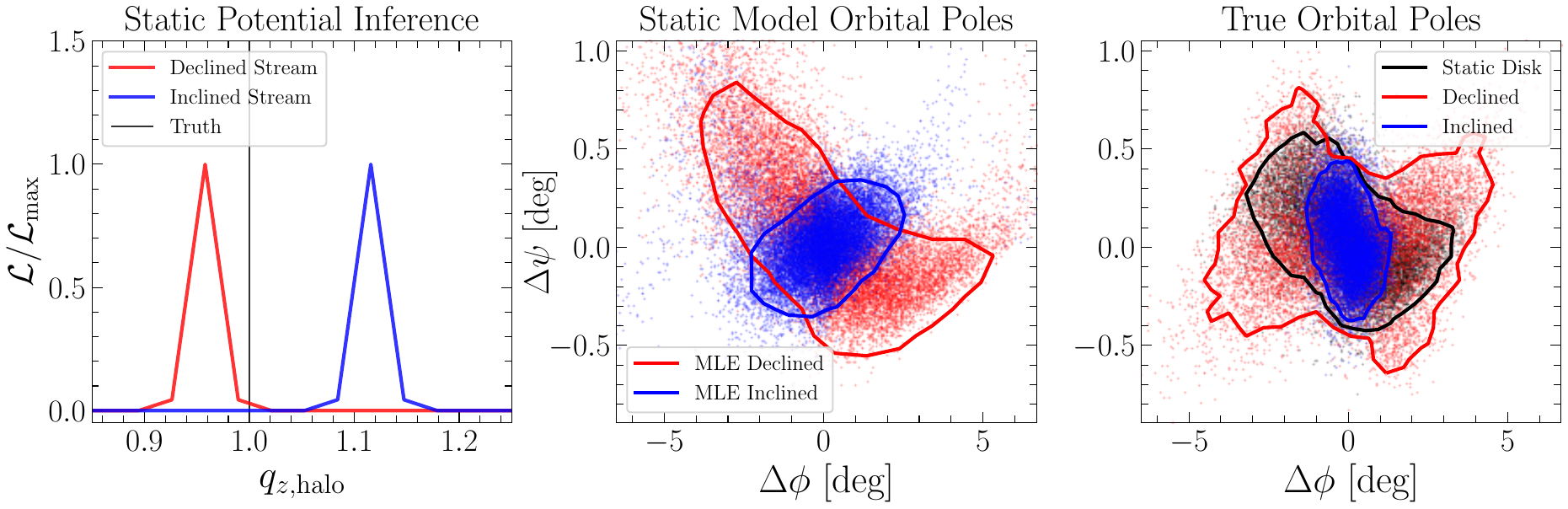}
    \caption{
    Constraints on the flattening of the dark matter halo estimated from streams can be biased when the disk's tilt is unaccounted for. The direction of bias can vary for each stream, depending on whether the progenitor has experienced inclining or declining of its orbital inclination due to the disk's time-dependence. \textbf{Left Panel:} Static potential likelihood for two streams, as a function of the dark matter halo's flattening. The red curve is for a stream whose orbital inclination was lowered by the disk's tilt in the past, while the blue curve corresponds to a stream whose orbital inclination was increased due to the disk's tilt in the past. All streams share the same final progenitor location, and the true halo is spherical. \textbf{Middle Panel:} The orbital pole distribution for the best-fit streams from the left panel; the static fit to the inclined stream (blue) has undergone less differential precession compared to the declined stream (red). \textbf{Right Panel:} The true orbital pole distribution of the declined and inclined streams (red and blue, respectively), and a stream generated in a spherical halo with a static disk (black).}
    \label{fig: Inference_test}
\end{figure*}

While a robust detection of a tilting disk would provide a novel constraint on the evolution of the Milky Way predicted by $\Lambda\rm{CDM}$, it is also important to consider whether such time-dependence can bias existing constraints on the potential. For instance, there is not a clear consensus on the shape of the Milky Way's dark matter halo. Some works suggest a prolate halo, where others suggest an oblate halo even when applying the same modeling framework to different streams (e.g., \citealt{2016ApJ...833...31B,2021MNRAS.502.4170R}). It is therefore necessary to consider how failure to account for time-dependence might bias these constraints, especially as the data become more informative thanks to high-precision astrometric surveys like {\it Gaia}.

As a simple experiment, we take the two streams presented in Fig.~\ref{fig: HeatingCoolingStream} as ``observations" and perform a mock inference of the potential using the full 6D phase-space information for both streams. These streams share the same final progenitor location, but one experienced the declining effect due to the disk's tilt, while the other experienced the inclining effect. The  tilting rate in both cases is $15~\rm{deg}/\rm{Gyr}$. 

In our mock inference, we will incorrectly assume a static potential and fix the parameters of the disk to their true values and orientation at the present day. For the halo potential, we use a NFW halo with flattening $q_{z,\rm{halo}}$ in the density \citep{2003ApJ...585..151L}. In reality, the streams were generated in a spherical halo, though we explore to what extent our assumption of a static potential will bias constraints on the flattening. To carry out a mock inference, we need a likelihood to connect a given potential model to the data. To construct this likelihood, we follow the approach outlined in \citet{2014ApJ...795...94B} and center a series of Gaussians on each of the particles of the observed streams. The likelihood of a new phase-space point is then the sum of the Gaussian distributions evaluated at the new point. 
We choose the standard deviation of each Gaussian to provide a reasonably smooth likelihood, determined using a cross-validation bandwidth selection. For additional details and motivation regarding the likelihood, we refer readers to \citet{2014ApJ...795...94B}.  

With a likelihood in hand, we generate a grid of streams with the final phase-space location of the progenitor fixed to the true value but in halos with different z-axis flattening values. For each simulation, we evaluate the likelihood as a product over all stream particles. The left panel of Fig.~\ref{fig: Inference_test} shows the likelihood curves for our mock inference. The red curve is the likelihood function when assuming that the declined stream represents the observations, while the blue curve corresponds to the case of the inclined stream. The true halo flattening is $q_{z,\rm{{halo}}}=1$.

From Fig.~\ref{fig: Inference_test}, we find that a tilting disk can bias an inference of the halo's flattening at the 5--10\% level. We expect that the magnitude of the bias is dependent on the present location and orbital history of the stream, though we defer an exploration of this dependence to future work. Most concerning is the tendency for the two streams with the same progenitor to have bias in opposite directions. Namely, the declined stream (red) prefers an oblate halo with $q_{z,\rm{halo}} \approx 0.95$, while the inclined stream prefers a prolate halo with $q_{z,{\rm halo}} \approx 1.1$. When applying the same test to different streams, we find similar results. The morphology and kinematics of a declined stream are different enough from its inclined counterpart to produce a biased constraint on the potential with an opposite sign. The magnitude of the bias is less for the declined stream than the inclined stream by around $5\%$. This does not appear to be true in general, but rather depends on the specific orbit of the stream considered and (e.g.) the number of pericenters it experiences during integration.

We can understand why the inclined and declined streams bias our inference of the potential in opposite directions  through the discussion in \S\ref{sec: effect_stream_width}. Namely, when a stream undergoes the inclining effect in the past, its differential precession frequency, and therefore azimuthal torque distribution, is suppressed. Conversely, a declining stream experiences a larger amount of differential precession and a wider amount of dispersion in azimuthal torques. This is reflected in Fig.~\ref{fig: Differential_Torque} (bottom panel). When trying to infer an incorrectly static potential, this means that the maximum likelihood estimate (MLE) for $q_{z,\rm{halo}}$ will tend to prefer a value that leads to a smaller azimuthal torque dispersion for the inclined stream and a larger azimuthal torque dispersion for the declined stream. 

We illustrate how a static potential can give rise to distinct torque dispersions in the middle panel of Fig.~\ref{fig: Inference_test}, where we plot the orbital pole distribution of the best-fit stream models from the left panel, which were fit under incorrect assumptions (i.e., a flattened halo with a static disk). The true pole vector distribution for both streams are shown in the right panel. The black distribution represents the pole vectors for a stream generated in the fiducial static potential with a spherical halo. As expected, the MLE for the inclined stream is less horizontally extended than the declined stream. The azimuthal extent of both mock streams mirror the ``observed" streams in their angular momentum footprint, shown in the right panel.

The result of this discussion is that a tilting disk can bias potential constraints on the halo flattening at the 5--10\% level, though the direction of the bias can differ depending on whether the stream experienced the declining effect or the inclining effect due to the disk's time-dependence. This level of bias, and the direction, is similar to constraints on the halo's flattening from Pal 5 and GD-1 in \citet{2016ApJ...833...31B}. In their work, Pal 5 prefers a z-axis flattening of $q_{\rm z,\rm{halo}} \approx 0.9$, while GD-1 prefers a more oblate solution of $q_{\rm z,\rm{halo}} \approx 1.3$ at the best fit. \citet{2021MNRAS.502.4170R} finds a similar result for Pal 5 and GD-1, and also for the Orphan--Chenab and Helmi streams. Especially for more local streams like Pal 5 and GD-1, this tension could be relaxed by determining if a consistent tilting disk experienced by both streams reduces the disagreement in $q_{z,\rm{halo}}$. We leave a test of this possible solution to future work.

\section{Discussion}\label{sec: discussion}

\subsection{Caveats}\label{sec: future}
We have made several idealizations that could be improved upon in future work. We highlight these below.

First, we treat the disk as a rigid body. In reality, the response of a stellar disk to (e.g.) satellite mergers could cause more of a warping effect than a global tilt. However, in a series of zoom-in cosmological simulations, \citet{2023MNRAS.518.2870D} finds that the global morphology of the disk does not change substantially in response to an infalling satellite. The most significant difference is due to vertical heating, which can induce up to a 30\% increase in the disk's scale-height. Similarly, other works have argued that the inner disk behaves rigidly in response to infalling satellites \citep{1986MNRAS.218..743B,1998ApJ...506..590S,2006MNRAS.370....2S,2023MNRAS.518.2870D}. Future work could consider a more realistic treatment of the disk.

Second, we have considered sustained values and directions for $\mathbf{\Omega}_{\rm Disk}$, while the realistic behavior is for the angular momentum vector of the disk to precess (see, e.g., the Briggs diagrams in \citealt{2023MNRAS.518.2870D}). If disk tilting is caused by satellite mergers, precession of the disk is a natural outcome since there is not necessarily an alignment between the infalling satellite's angular momentum and that of the disk. We can be certain that aligning the tilting axis with the normal to the disk leads to a smaller effect on streams. This behavior is described in Appendix~\ref{app: arb_tilting_axis} and is unsurprising since an axisymmetric disk is invariant to rotations about its symmetry axis. 

Third, we have neglected other time-dependent perturbations like a rotating bar or the currently infalling Large Magellanic Cloud. For streams with closer-in apocenters like Pal 5, we expect the bar to play a substantial role in shaping the stream's morphology (see, e.g., \citealt{2016MNRAS.460..497H,2017MNRAS.470...60E,2017NatAs...1..633P,2020ApJ...889...70B}). However, the dominant effect of the bar is typically localized to the regions of a stream which suffered a close encounter. Disk tilting produces a more global effect on tidal tails and could in principle be disentangled from the more local effect of the bar. 

Fourth, there are also resonant effects due to a tilting disk that could have an appreciable impact on certain streams. The timescale of orbital plane precession is similar to the expected disk-tilting rates (Fig.~\ref{fig: RadializeCircularizeDemo}), and in some of our simulations, we have found this can lead to chaotic orbits that would otherwise be regular in a static potential. We have not explored the ubiquity or observational signatures of such resonances in a tilting disk potential. 

Fifth, we utilize a spherical dark matter halo in our simulations, though an additional axisymmetric component introduces another source of torque in addition to a tilting disk. Because potentials obey superposition, we do not expect the inclusion of a flattened halo to alter our conclusions. Observationally, we discuss degeneracies between a tilting disk and flattened dark matter halo in \S\ref{sec: implications}.

\subsection{Prospects for Detecting the Disk's Angular Momentum Evolution using Streams}\label{sec: prospections}
We have shown that stellar streams can provide a probe of the disk's angular momentum evolution. For realistic disk tilting rates of $\sim 10$--$15~\rm{deg}/\rm{Gyr}$, the dominant effect is to shift the on-sky track and width of the stream. While a time-dependent disk potential can, unsurprisingly, produce more diffuse streams, we have shown that it can also produce more narrow streams. This is distinct from other time-dependent perturbations in the Milky Way such as the galactic bar or spiral arms, which will typically have the effect of producing kinematically hotter, broader streams (e.g., Fig.~4 in \citealt{2020ApJ...889...70B}).

When full 6D phase-space information is available, the orbital pole vector distribution of a stream is sensitive to the disk's time dependence, especially for streams with close-in apocenters. For these streams, one can expect a systematic tilt in the pole vector distribution that is not otherwise captured in a static disk model with a spherical halo. Provided that the disk's angular momentum evolution is gradual, even the spatial track of the stream exhibits a similar tilt and can be used independent of velocity information to constrain $\mathbf{\Omega}_{\rm Disk}$. This tilting effect indicates the presence of vertical torques, or that $L_z$ is not conserved. 

Of course, other aspects of the potential could break the conservation of $L_z$, such as a non-axisymmetric mass distribution. However, as discussed in \S\ref{sec: implications}, a tilting disk can bias constraints on shape parameters of the dark matter halo when the disk's time-dependence is unaccounted for. Constraints on halo flattening are biased at the 5--10\% level when the potential is assumed to be static, though the direction of the bias (oblate versus prolate) varies on a per-stream basis. This is because inclining or declining streams experience distinct torque dispersions throughout their lifetimes (Fig.~\ref{fig: Differential_Torque}), mimicking the effect of an incorrectly oblate or prolate halo. Thus, if the disk has undergone signicant angular momentum evolution throughout its lifetime, it may be difficult to posit a single, static global potential model that simultaneously reproduces the pole vector distribution of multiple streams. If the pole distribution of multiple streams with relatively close-in apocenters can be accurately modeled with a single tilting disk and a simple halo component, this would provide significant evidence for the disk's angular momentum evolution. Otherwise, a single stream alone cannot be used to definitively infer the angular momentum evolution of the disk. In Appendix~\ref{app: arb_tilting_axis}, we provide a more quantitative argument for this statement.

Because of degeneracies between the pole distribution of a stream and the shape of (e.g.) the halo potential, we argue that modeling the width of multiple streams simultaneously should provide a more robust constraint on the disk's time-dependence. From Fig.~\ref{fig: 25percent_change}, we expect that several inclined streams in the inner halo should be sensitive to both stream narrowing and fanning. For a given stream, the orientation between the disk's tilting axis and the progenitor's angular momentum vector largely determines which effect will occur. For a given choice of $\mathbf{\Omega}_{\rm Disk}$, the fanning or narrowing effect is predictable and can be used to test different tilting scenarios. 

If several streams are found to be systematically more diffuse or narrow than a static potential model would otherwise predict, it is worth considering if the population of stream widths and on-sky tracks could be reproduced in a simple global potential with a time-dependent disk component. For instance, we find a strong match to Pal 5's morphology and width in Fig.~\ref{fig: Pal5} when assuming $\Omega_{\rm Disk} = -15~\rm{deg}/\rm{Gyr}$ with rotation about an axis in the midplane of the disk. If the width of several other streams are better characterized by this simple model, this would provide a more robust constraint on $\mathbf{\Omega}_{\rm Disk}$.

Still, streams should not provide the only line of evidence for the disk's angular momentum evolution. There are also prospects for a more direct detection using distant Quasars and {\it Gaia} astrometry \citep{2014ApJ...789..166P}. \citet{2014ApJ...789..166P} predict that {\it Gaia} should be able to detect a tilting disk with a frequency of at least $0.28~\rm{deg}/\rm{Gyr}$ using this approach. However, their method requires the disk to be tilting currently. Streams, as we have shown, are sensitive to the time-evolution of the disk's angular momentum. If the disk underwent a large transition over a few Gyr and then settled (e.g., \citealt{2023arXiv231013050C}), globular cluster streams should retain a memory of this transition and provide a testbed for dynamically constraining the formation of the stellar disk.

\section{Summary and Conclusion}\label{sec: conclude}
Stellar disk tilting is an expected outcome of hierarchical assembly in $\Lambda\rm{CDM}$. In this paper, we have studied the effect of realistic disk tilting rates on the dynamics and morphology of stellar streams. The main results of our work are summarized below:

\begin{itemize}
    \item \textbf{Track offsets:} Streams in the inner halo ($r_{\rm apo} \lesssim 20~\rm{kpc}$) provide a sensitive probe of disk tilting, with shifts in their on-sky $\phi_1-\phi_2$ tracks at the level of $\sim 1~\rm{deg}$. This is detectable, especially as dynamical models of the Galaxy become more precise. 

    \item \textbf{Stream Width:} Disk tilting can produce (unsurprisingly) more diffuse streams, though it can also produce more narrow streams. The divergence of stream widths is predictable for a given $\mathbf{\Omega}_{\rm Disk}$, with the direction of the disk's tilt relative to the stream progenitor's angular momentum vector determining which streams should fan out versus narrow. Studying the widths of multiple streams simultaneously provides a robust test of disk tilting, especially if the widths of multiple streams can be recovered under a single tilting disk model.

    \item \textbf{Pal 5's Tidal Tails:} The on-sky track and width of Pal 5 is well-captured in a tilting disk potential, with $\Omega_{\rm Disk} = 15~\rm{deg}/\rm{Gyr}$. The direction of the disk's tilt is such that Pal 5 would have been on a more inclined orbit in the past. The other tilting direction lowers the inclination of Pal 5's orbit in the past and is ruled out by the morphology of Pal 5's tidal tails.

    \item \textbf{Slanted Orbital Poles and Potential Reconstruction Bias:} For streams in the inner halo, a tilting disk breaks the conservation of $L_z$, leading to vertical torques and a slanted stream orbital pole distribution. If the disk's tilt is not accounted for, these vertical torques can bias constraints on the flattening of the inner dark matter halo. Depending on the relative orientation between the disk's tilt and the stream progenitor's angular momentum, some streams will bias constraints towards an oblate halo, while others will prefer a prolate halo. Meanwhile, the true halo potential in our simulations is spherical. The magnitude and direction of the bias is similar to differing constraints on the real Milky Way halo's flattening derived from multiple streams.

\end{itemize}

In total, the kinematics and morphology of stellar streams are imprinted with observable signatures due to disk tilting. This provides a promising avenue to constrain $\mathbf{\Omega}_{\rm Disk}$ in the Milky Way, using a population of stellar streams. On the other hand, the sensitivity of nearby streams to the disk's angular momentum evolution over the past few Gyrs challenges the validity of static models for the Galaxy and the inner dark matter halo. This motivates the development of time-dependent models for the Milk Way that capture its merger history and global response to external perturbations.

\section*{Acknowledgements}
JN is supported by a National Science Foundation Graduate Research
Fellowship, Grant No. DGE-2039656. Any opinions, findings, and conclusions
or recommendations expressed in this material are those of the author(s) and
do not necessarily reflect the views of the National Science Foundation.  ML is supported by the Department of Energy~(DOE) under Award Number DE-SC0007968 and the Simons Investigator in Physics Award. DE acknowledge support through
ARC DP210100855.

\software{Numpy \citep{harris2020array}, Matplotlib \citep{Hunter:2007}, SciPy \citep{2020SciPy-NMeth}, Gala \citep{adrian_price_whelan_2020_4159870}, Astropy \citep{astropy:2018} }

\appendix
\section{Equations of Motion in Co-Rotating Frame}\label{app: EqnsofMotion}
Throughout the paper, we assume a simplified scenario where the disk tilts around an axis in its plane. We use a cylindrical coordinate system consisting of $(R,\phi,z)$, where the z-axis is perpendicular to the disk's mid-plane. In a coordinate system co-rotating with the disk with frequency $\mathbf{\Omega} = \Omega \hat{\mathbf{x}}$, the Hamiltonian is
\begin{equation}
\begin{split}
    H_J &= H - \mathbf{\Omega} \cdot \mathbf{L} \\
    &= \frac{1}{2}\left[p^2_R + \frac{p^2_\phi}{R^2} + p^2_z\right] + \Phi(R,z) \\ &+ \Omega \left[z\frac{p_\phi }{R}\cos\phi + \left(zp_R - Rp_z\right)\sin\phi \right],
\end{split}
\end{equation}
where $H$ represents the usual Hamiltonian in an inertial frame and $H_J$ is the Hamiltonian in the rotating frame, also called the Jacobi Integral. A feature of rotating coordinate systems is that the cartesian conjugate momentum in the rotating frame, $\mathbf{p}$, is equal to the cartesian momentum in the inertial frame.  That is, the conjugate momentum (per unit mass) in the co-rotating frame is 
\begin{equation}
    \mathbf{p} = \mathbf{\dot{x}}_{\rm corot} + \mathbf{\Omega} \times \mathbf{x}_{\rm corot} = \mathbf{\dot{x}}_{\rm inertial}.
\end{equation}
Thus, $(p_R, p_\phi, p_z)$ can be viewed as the usual cylindrical momentum components in an inertial frame (i.e., $\dot{R}, R^2 \dot{\phi}, \dot{z}$) but projected along the time-dependent axes of the rotating cylindrical coordinate system defined by $\hat{\mathbf{R}}(t), \hat{\mathbf{\phi}}(t), \hat{\mathbf{z}}(t)$. With the Hamiltonian and its terms defined, the time derivatives for the conjugate momenta can be obtained via Hamilton's equations. The result is
\begin{equation}
    \begin{split}
        \dot{p_R} &= -\frac{\partial H_J}{\partial R} =\frac{p^2_\phi}{R^3} -\frac{\partial \Phi}{\partial R} + \Omega\left[-2z\frac{p_\phi}{R^2}\cos\phi - p_z\sin\phi \right] \\
        \dot{p}_\phi &=-\frac{\partial H_J}{\partial \phi} = \Omega\left[z\frac{p_\phi}{R}\sin\phi - \left(zp_R - Rp_z \right)\cos\phi\right] \\
        \dot{p}_z  &= -\frac{\partial H_J}{\partial z} = -\frac{\partial \Phi}{\partial z}+ \Omega\left[\frac{p_\phi}{R}\cos\phi  + p_R\sin\phi\right].
    \end{split}
\end{equation}
The time-derivatives for position in the co-rotating cylindrical coordinate system are
\begin{equation}
    \begin{split}
        \dot{R} &= \frac{\partial H_J}{\partial p_R} = p_R + \Omega z\sin\phi  \\
        \dot{\phi} &= \frac{\partial H_J}{\partial p_\phi} = \frac{p_\phi}{R^2} + \Omega \frac{z}{R}\cos\phi \\
        \dot{z} &= \frac{\partial H_J}{\partial p_z} = p_z - \Omega R\sin\phi.
    \end{split}
\end{equation}

The equations of motion in a static axisymmetric potential are a special case, with $\Omega = 0$. For the static case, $p_\phi$ is an integral of motion, representing the z-component of angular momentum. In the rotating case, $\dot{p}_\phi$ is no longer conserved and instead varies non-trivially with position and velocity. We can see already that a key signature of a tilting axisymmetric mass component is excess precession of the orbital plane, such that none of the angular momentum components are individually conserved. 

Identifying $p_\phi$ with $L_z$ (where $z$ refer's to the co-rotating cylindrical z-coordinate), the evolution of the z-component of the angular momentum is
\begin{small}
\begin{equation}
\begin{split}
        \tau_z &= \frac{dL_z}{dt} \\
        &= \Omega \left[\left(R\dot{z} - z\dot{R}\right)\cos{\phi}  + \left(zR\dot{\phi} + \Omega R^2\cos\phi \right)\sin\phi\right].
\end{split}
\end{equation}
\end{small}

\section{Small Vertical Oscillations in the Rotating Frame}\label{sec: small_vertical_osc}
In this section, we take $\Phi$ to be a a general flattened axisymmetric potential, which is a function of the flattened radius $r = \sqrt{x^2 + y^2 + z^2/q^2}$. We highlight the vertical behavior of orbits in a tilting axisymmetric potential when the amplitude of oscillation is small.

We assume that the potential $\Phi$ is rotating with frequency $\mathbf{\Omega} = \Omega \hat{\mathbf{x}}$. In the non-inertial frame, co-rotating with this frequency, $\Phi$ is static and the vertical equation of motion is
\begin{equation}
     \ddot{z} = -\Phi^\prime\left(r\right)\frac{z}{q^2 r} - \Omega^2  z.
\end{equation}
Expanding this expression to first order in $z = \Delta z$,
\begin{equation}\label{eq: Delta_z_dot_dot}
\begin{split}
    \Delta \ddot{z} &= -\Phi^\prime(r_0) \frac{\Delta z}{q^2 r_0} - \Omega^2 \Delta z \\
    &= -\left(\frac{\Omega_{\rm Orbit}^2}{q^2} + \Omega^2 \right)\Delta z.
\end{split}
\end{equation}
The solution to Eq.~\ref{eq: Delta_z_dot_dot} is
\begin{equation}
    \Delta z(t) = \Delta z_0 \cos\left[ \left(\frac{\Omega_{\rm Orbit}^2}{q^2} + \Omega^2\right) t + \phi_0 \right].
\end{equation}
Thus, in the co-rotating frame, an orbit with small oscillations around the mid-plane of the flattened potential will execute oscillatory vertical motion just as it would in the static case, though with a greater frequency of oscillation. For small vertical oscillations, the time-averaged $\langle \Delta z(t) \rangle = 0$. Because we have worked in a frame rotating with the potential, this implies that the orbital plane will tend to co-rotate with the potential so that the angular momentum of the orbit averages out to point along the $z-$axis.

\begin{figure}
    \centering
    \includegraphics[scale=0.5]{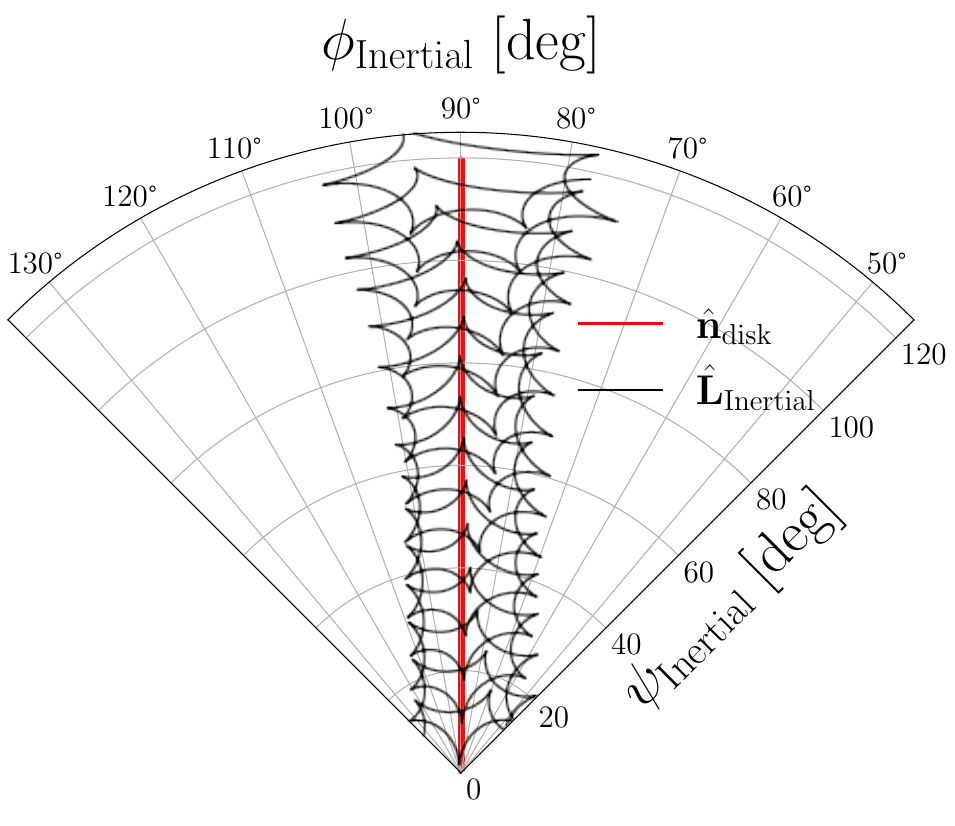}
    \caption{Orbital pole diagram in an inertial frame. The red line depicts the time-evolution of the tilting disk's normal vector, $\hat{\mathbf{n}}_{\rm disk}$. The disk tilts around $\hat{\mathbf{x}}$. $\psi_{\rm intertial}$ is a spherical angle measured from the inertial z-axis, and $\phi_{\rm inertial}$ is the azimuthal polar angle in the inertial frame. The angular momentum pole vector of an orbit initialized with small vertical oscillations about the disk's midplane is shown in black. The orbital plane gradually adjust to remain aligned with the disk. }
    \label{fig: OrbitAngularMomentumTrack}
\end{figure}

We illustrate this behavior in Fig.~\ref{fig: OrbitAngularMomentumTrack}, which shows the angular momentum unit vector of an orbit in the tilting disk potential, as viewed in an inertial frame (black). The orbit has a typical vertical extent of $\Delta z_0 \approx 2~\rm{kpc}$. The time-evolution of the disk's normal vector, $\hat{\mathbf{n}}_{\rm disk}$, is shown as the red line. In the inertial frame, we can see that even for a moderately large value of $\Delta z_0$,  the angular momentum of the orbit tends to track with the disk's $z-$axis. That is, the orbital plane adjusts to remain aligned with the disk in a time-averaged sense. 

\section{Effect of an Arbitrary Tilting Axis and the Importance of Multiple Streams}\label{app: arb_tilting_axis}
If the disk tilts around an arbitrary axis $\hat{\mathbf{n}}_{\rm tilt}$ by some increment $\Delta \Psi$ over a lookback time interval $\Delta t_{\rm back} > 0$, a particle with the initial precession or nutation frequency $\Omega_0$ and orbital inclination $\psi_0$ will acquire the new frequency
\begin{equation}\label{eq: general_omega_expression}
    \Omega_{\rm new} \approx  \Omega_0   + \frac{d\Omega}{d\psi}\Big\vert_{\psi_0} \frac{d\psi}{d\Psi}\frac{d\Psi}{dt_{\rm back}}\Delta t_{\rm back},
\end{equation}
where $\frac{d\Psi}{dt_{\rm back}} \equiv \Omega_{\rm Disk}$. The term $d\psi/d\Psi$ describes how the orbital inclination changes due to a rotation $\Delta \Psi$ about the tilting axis, $\hat{\mathbf{n}}_{\rm tilt}$. If $\mathbf{L}$ is orthogonal to $\hat{\mathbf{n}}_{\rm tilt}$, then $d\psi/d\Psi = \pm 1$. If the pair are parallel, then the derivative is zero, since rotation about the angular momentum vector's axis does not change the orbital inclination.

The term $d\psi/d\Psi$ can be obtained using the vector expression for rotational velocity (i.e., $\mathbf{\dot{\mathbf{r}}} = \boldsymbol{\omega} \times \mathbf{r}$), specialized to the case of angular momentum and its first derivative (torque). From this, the relation between $\psi$ and $\Psi$ is
\begin{equation}
    \frac{d\psi}{d\Psi} = \hat{\boldsymbol{\phi}}\left(\mathbf{L}\right) \cdot \hat{\mathbf{n}}_{\rm tilt},
\end{equation}
which can be substituted into Eq.~\ref{eq: general_omega_expression} to obtain
\begin{equation}
    \Omega_{\rm new} \approx  \Omega_0   + \frac{d\Omega}{d\psi}\Big\vert_{\psi_0} 
    \left[\hat{\boldsymbol{\phi}}\left(\mathbf{L}_0\right) \cdot \hat{\mathbf{n}}_{\rm tilt}\right]
    \frac{d\Psi}{dt_{\rm back}}\Delta t_{\rm back},
\end{equation}
where $\mathbf{L}_0$ is the orbital angular momentum today. From this expression, the change in the angular momentum precession or nutation frequencies depends on both local properties of the potential ($d\Omega/d\psi$), the tilting rate of the disk ($d\Psi/dt_{\rm back}$), and the geometry of the rotation (bracketed term). The bracketed term is maximized when the tilting axis is parallel to $\hat{\boldsymbol{\phi}}$, and minimized when orthogonal to $\hat{\boldsymbol{\phi}}$. This is unsurprising upon inspection of Fig.~\ref{fig: coord_sys}. 

For an intermediate tilting axis, there is an effective disk tilting frequency of 
\begin{equation}\label{eq: effective_tilt}
    \mathbf{\Omega}_{\rm Disk, eff} \equiv \left[\hat{\boldsymbol{\phi}}\left(\mathbf{L}_0\right) \cdot \hat{\mathbf{n}}_{\rm tilt}\right]
    \frac{d\Psi}{dt_{\rm back}}.
\end{equation}
The implication of the effective tilting frequency is a clear degeneracy between (a) the tilting axis of the disk and (b) the magnitude of the disk's tilting rate. For instance, the effect of a $10~\rm{deg}/\rm{Gyr}$ tilt can be achieved by either tilting the disk at this rate around an axis in its midplane ($\beta = 90~\rm{deg}$ in Fig.~\ref{fig: coord_sys}), or tilting the disk at a higher rate around some intermediate axis ($\beta < 90~\rm{deg}$). Both can yield the same effective frequency in Eq.~\ref{eq: effective_tilt}. We therefore expect that our choice throughout the paper of considering a tilting axis in the midplane of the disk does not greatly narrow the scope of our results. Provided that we test different values of $\Omega_{\rm Disk}$ for a given stream, the effect of varying the inclination of the tilting axis should produce similar results. 

For constraining $\mathbf{\Omega}_{\rm Disk}$, Eq.~\ref{eq: effective_tilt} highlights the necessity of using multiple streams at once. For a single stream, we can only constrain the effective tilting rate, though, because different streams will have different values of $\mathbf{L}_0$, $\mathbf{\Omega}_{\rm Disk, eff}$ will also be different. The intrinsic disk tilting rate, $\hat{\mathbf{n}}_{\rm tilt} (d\Psi/dt_{\rm back})$, can only be inferred by stitching together constraints on the effective tilting rate from multiple streams simultaneously. 

\subsection{Degeneracies in a Time-Dependent Potential}
More broadly, it is interesting to consider whether introducing a flexible time-dependent component to the potential will always introduce similar degeneracies to Eq.~\ref{eq: effective_tilt}. In a simple static potential, if one acceleration measurement can be reliably inferred then the parameters of the model have been determined (assuming a one-to-one relation between model parameters and the acceleration at a point). In a more flexible time-dependent potential incorporating several tracers might be a necessity, since this one-to-one mapping may no longer hold due to degeneracies like Eq.~\ref{eq: effective_tilt} (i.e., different orbital geometries, tilting axes, and intrinsic tilting rates giving the same \emph{effective} tilting rate). A tilting disk provides one such clear example, though it will be important to consider the ubiquity of degeneracies like Eq.~\ref{eq: effective_tilt} in attempting to constrain other time-dependent features of galaxies.

\bibliography{thebib}
\end{document}